\documentclass[usenatbib]{mnras}
\usepackage{epsfig}
\usepackage{psfig}
\usepackage{amsmath,amssymb}
\usepackage{txfonts}
\allowdisplaybreaks
\def\FVFPS{{\sc fvfps\,}}

\renewcommand\d{{\rm d}}

\newcommand\Ezero{{E_0}}
\newcommand\Eu{{E_{\rm u}}}
\newcommand\Etwob{{E_{\rm 2b}}}

\newcommand\Lzero{L_0}
\newcommand\Lcirc{{L_{\rm circ}}}
\newcommand\Lmax{{L_{\rm max}}}
\newcommand\Lu{{L_{\rm u}}}
\renewcommand\i{{\rm i}}
\newcommand\kms{{\rm \,km\,s^{-1}}}
\newcommand\kpc{{\rm \,kpc}}

\newcommand\M{M}

\newcommand\Mtwob{{\M}_{\rm 2b}}
\newcommand\Mtot{{\M}_{\rm tot}}
\newcommand\Mcen{{\M}_{\rm cen}}

\newcommand\Msat{\M_{\rm sat}}
\newcommand\musat{\mu_{\rm sat}}

\newcommand\Msun{{\M}_{\odot}}

\newcommand\Mref{\M_{\rm ref}}

\newcommand\rcen{r_{\rm cen}}
\newcommand\rapo{r_{\rm apo}}

\newcommand\rs{r_{\rm s}}
\newcommand\rperi{r_{\rm peri}}
\newcommand\rperitwob{r_{\rm peri,2b}}
\newcommand\ra{r_{\rm a}}

\newcommand\rvir{{r_{200}}}
\newcommand\rt{{r_{\rm t}}}

\newcommand\tu{t_{\rm u}}

\newcommand\tfin{t_{\rm fin}}
\newcommand\tmax{t_{\rm max}}

\newcommand\varepsilonDM{{\varepsilon_{\rm DM}}}
\newcommand\varepsilonsat{{\varepsilon_{\rm sat}}}

\newcommand\vapo{{v_{\rm apo}}}
\newcommand\vz{{v_z}}
\newcommand\vx{{v_x}}
\newcommand\vy{{v_y}}
\newcommand\vr{{v_r}}

\newcommand\vcirc{v_{\rm circ}}
\newcommand\vtan{v_{\rm tan}}

\newcommand\vu{v_{\rm u}}

\newcommand\rhoa{\rho_{\rm a}}

\newcommand\sigmacen{\sigma_{\rm cen}}

\renewcommand\S{\mathcal{S}}

\newcommand\yr{{\rm \,yr}}
\begin{document}
%\begin{landscape}

\date{Accepted 2017 January 12. Received 2017 January 6; in original form 2016 November 3.}

\title[Growth of central galaxies]{The special growth history of central galaxies in groups and clusters}

\author[C. Nipoti]{Carlo Nipoti\thanks{E-mail: carlo.nipoti@unibo.it}
  \\Department of Physics and Astronomy, Bologna University,
  viale Berti-Pichat 6/2, I-40127 Bologna, Italy }

% $^{1}$

\maketitle
\begin{abstract}
Central galaxies (CGs) in galaxy groups and clusters are believed to
form and assemble a good portion of their stellar mass at early times,
but they also accrete significant mass at late times via galactic
cannibalism, that is merging with companion group or cluster galaxies
that experience dynamical friction against the common host dark-matter
halo. The effect of these mergers on the structure and kinematics of
the CG depends not only on the properties of the accreted satellites,
but also on the orbital parameters of the encounters.  Here we present
the results of numerical simulations aimed at estimating the
distribution of merging orbital parameters of satellites cannibalized
by the CGs in groups and clusters. As a consequence of dynamical
friction, the satellites' orbits evolve losing energy and angular
momentum, with no clear trend towards orbit circularization.  The
distributions of the orbital parameters of the central-satellite
encounters are markedly different from the distributions found for
halo-halo mergers in cosmological simulations. The orbits of
satellites accreted by the CGs are on average less bound and less
eccentric than those of cosmological halo-halo encounters. We provide
fits to the distributions of the central-satellite merging orbital
parameters that can be used to study the merger-driven evolution of
the scaling relations of CGs.
\end{abstract}

\section{Introduction}

Central galaxies (CGs) in galaxy groups and clusters are typically
massive early-type galaxies (ETGs) with relatively old stellar
populations and little ongoing star formation. In the currently
favoured formation model \citep{Mer85,Tre90,Dub98,Rus09}, CGs form and
assemble a good portion of their stellar mass at relatively early
times (say $z\gtrsim 1$), during the virialization of their host
systems. At later times (say $z\lesssim 1$) further growth of the CGs
occurs, when their hosts are almost virialized, through galactic
cannibalism, that is the accretion of companion galaxies orbiting in
the common host dark-matter (DM) halo.  Galactic cannibalism, which
was originally proposed as the dominant process to form CGs in
clusters \citep{Ost75,Whi76,Hau78}, is driven by dynamical friction
\citep{Cha43}, so the CGs must accrete preferentially relatively
massive satellite galaxies \citep{Ost75}, because the
dynamical-friction timescale is inversely proportional to the mass of
the decelerated object.  As these accreted satellite galaxies live in
the dense environment of the group or cluster core, they are expected
to be poor in cold gas and not to have significant ongoing star
formation. For these reasons, the growth of CGs by galactic
cannibalism should be well represented by a sequence of
dissipationless (dry) mergers. This is consistent with the
observational finding that CGs are invariably ETGs. The cumulative
effect of these accretion episodes is important in the build-up of
these massive galaxies: both observational
\citep{Mar14,Bel16,Buc16,Vul16b} and theoretical
\citep{Del07,Ton12,Sha15} studies suggest that the stellar mass of CGs
in clusters of galaxies has grown by up to a factor of two since
$z\approx 1$. Similar growth rates are found in simulations of group
CG formation \citep{Fel10}.

CGs are observed to follow relatively tight empirical scaling laws,
relating stellar mass (or luminosity), size and stellar velocity
dispersion \citep{Ber07,Liu08,Vul14}. There are indications that these
scaling relations are even tighter than those of normal ETGs
\citep{Ber07,Mon16}.  If the idea that CGs grow significantly via
galactic cannibalism is correct, the high-mass end of the ETG scaling
relations must be built and maintained by dry mergers
\citep{Nip03b,Ber11}, which are known to have an impact on the scaling
laws \citep[][]{Nip03a,Boy05}.  The effect of the accretion of a
satellite on the size and velocity dispersion depends not only on the
properties of the satellite \citep[such as mass and velocity
  dispersion;][]{Naa09}, but also on the orbital parameters of the
encounter \citep{Boy06,Nip12}.  In cosmologically-motivated
simulations of formation of brightest group galaxies
\citep{Tar13,Tar15,Sol16,Per16}, the final CGs, though forming through
complex merger histories, are found to follow tight scaling relations,
similar to those observed for the real CGs. The knowledge of the
distribution of the merging orbital parameters could help understand
this finding.

In this paper we focus on the late cannibalism-driven growth of CGs
and we explore the properties of the central-satellite encounters.
The orbital parameters of these encounters are expected to be mainly
determined by the fact that each CG is located in a very special
place, that is the bottom of the deep gravitational potential well of
a group or cluster of galaxies. Mutatis mutandis, the process of the
accretion of satellites onto a CG is very similar to the process of
the formation of galactic stellar nuclei via accretion of galactic
globular clusters \citep{Gne97,Gne14}.  In essence, the idea is that
the satellite galaxies that end up being cannibalized by the CG not
only span a relatively small range in mass (they must be massive
enough for dynamical friction to be effective), but also have a
special distribution of the orbital parameters at the time of
encounter with the CG (because their orbits are determined mainly by
dynamical friction).  In the attempt to better characterize the
properties of the central-satellite encounters in groups and clusters,
in this work we follow with $N$-body simulations the evolution of the
orbital energy and angular momentum of model group and cluster
satellite galaxies under the effect of dynamical friction against the
cluster DM halo. These simulations allow us to determine the
distributions of the orbital parameters of the central-satellite
encounters. We then compare these distributions, relevant to the
growth of CGs in groups and clusters, to the distributions of the
orbital parameters measured for infalling satellite halos in
cosmological $N$-body simulations, which are expected to describe
accretion onto galaxies that are not centrals in either groups or
clusters.

The paper is organized as follows. The models and the $N$-body
simulations are described in Section~\ref{sec:model} and in
Section~\ref{sec:simu}, respectively. The results on the distributions
of the orbital parameters are presented in Section~\ref{sec:results}
and discussed in Section~\ref{sec:disc}. Section~\ref{sec:con}
concludes.

\section{Models and physical units}
\label{sec:model}

We study the growth of CGs with the help of idealized models. The
accretion of satellites occurs on timescales of the order of the
dynamical-friction timescale, which is longer than the
violent-relaxation timescale \citep{Lyn67}, that is the characteristic
time for the virialization of the host halo. Therefore, the satellite
galaxies accreted by the CGs are not expected to be directly traced by
the population of infalling cosmological sub-halos at their first
pericentric passage, but to belong to a population of satellites
roughly in equilibrium in the gravitational potential of the host halo
(see Section~\ref{sec:initial_dist} for a discussion).  Therefore, we
simply assume that in the initial conditions the halo is in
equilibrium and the satellite galaxies are on orbits extracted from
the halo distribution function. Of course, the distribution function
of the satellites must not necessarily be the same as that of the host
DM halo, but the two distributions are expected to be similar, because
they both originate from cosmological accretion. This is supported by
the analysis of halos in cosmological $N$-body simulations, in which
DM particles and sub-halos are found to have similar velocity
distribution \citep[][]{Gil04,Die04}.

Specifically, in our model the halo is represented by a spherical,
exponentially truncated \citet*[][NFW]{Nav96} density distribution
\begin{equation}
\rho (r)=\frac{\Mref}{r \left(r+\rs\right)^{2}}\exp\left[ -\left(\frac{r}{\rt}\right)^2\right],
\label{eq:rho}
\end{equation}
where $\rs$ is the scale radius, $\rt$ is the characteristic radius of
the exponential truncation and $\Mref$ is a reference mass.  We fix
$\rt=5\rs$, so the total DM mass is $\Mtot=4\pi\int_0^{\infty}\rho r^2
\d r\simeq0.82\Mref$.  The halo is assumed to have Osipkov-Merritt
\citep[][]{Osi79,Mer85} distribution function
\begin{equation}
f(Q)=\frac{1}{\sqrt{8}\pi^2}\frac{\d}{{\d} Q}\int_Q^0\frac{{\d}\rhoa}{{\d}\Phi}\frac{{\d}\Phi}{\sqrt{\Phi-Q}},
\label{eq:df}
\end{equation}
where $Q=E+{L^2}/{2\ra^2}$, $E$ is the specific energy, $L$ is the
specific angular-momentum modulus, and $\Phi$ is the gravitational
potential, such that
\begin{equation}
\nabla^2\Phi=4\pi G\rho,
\end{equation}
where $\rho$ is the density distribution~(\ref{eq:rho}). In
equation~(\ref{eq:df})
\begin{equation}
\rhoa(r)\equiv\left(1+\frac{r^2}{\ra^2}\right)\rho(r)
\end{equation}
is the so-called augmented density, where $\ra$ is the anisotropy
radius. The Osipkov-Merritt distribution function is such that the
velocity distribution is isotropic for $\ra/\rs=\infty$ and
increasingly radially anisotropic for smaller values of $\ra/\rs$. For
fixed $\ra/\rs$, the velocity distribution is isotropic in the
system's centre and more and more radially biased for increasing
distance from the centre.

For simplicity, the satellite is modeled as a ``rigid''
(i.e. represented by a single particle) system of mass $\Msat$
orbiting in the ``live'' (i.e. represented by an $N$-body system)
distribution of the halo (see Section~\ref{sec:tidal} for a discussion
of this assumption).  The CG is not modeled explicitly, but the
central mass distribution of the host can be thought of as due to the
CG. Therefore, we define as fiducial radius of the CG $\rcen=0.12\rs$
and as fiducial mass of the CG
\begin{equation}
\Mcen=\M(\rcen)\simeq0.0075\Mtot,
\label{eq:mcen}
\end{equation}
where $\M(r)=4\pi\int_0^r\rho r^2 \d r$ with $\rho(r)$ given by
equation~(\ref{eq:rho}).  The characteristic velocity dispersion of
the CG is $\sigmacen\equiv\sqrt{G\Mcen/\rcen}$. The specific choice of
the value of $\rcen/\rs$ is somewhat arbitrary: as long as the mass
ratio $\Mcen/\Mtot$ between the central galaxy and the host halo is
realistic, other values of $\rcen/\rs$ would be equally well
motivated. However, we verified that the main results of the present
work are not sensitive to the specific choice of $\rcen/\rs$, at least
for values in the range $0.1\lesssim \rcen/\rs\lesssim 0.15$
(corresponding to mass ratios $0.005\lesssim
\Mcen/\Mtot\lesssim0.011$).

The model can be rescaled by choosing the mass unit $\Mtot$ and the
length unit $\rs$. The corresponding units for other physical
  quantities are the velocity unit
\begin{equation}
\vu\equiv\sqrt{\frac{G\Mtot}{\rs}}\simeq 2074\left(\frac{\Mtot}{10^{14}\Msun}\right)^{1/2}\left(\frac{\rs}{100\kpc}\right)^{-1/2}\kms,
\label{eq:vu}
\end{equation}
the time unit
\begin{equation}
\tu\equiv\sqrt{\frac{\rs^3}{G\Mtot}}\simeq 4.72\times 10^7\left(\frac{\rs}{100\kpc}\right)^{3/2}\left(\frac{\Mtot}{10^{14}\Msun}\right)^{-1/2}\yr,
\label{eq:tu}
\end{equation}
the specific-energy unit $\Eu\equiv\vu^2$ and the specific
angular-momentum unit $\Lu\equiv\rs\vu$. It follows that, for
instance, for a system with $\Mtot=10^{14}\Msun$ and $\rs=200\kpc$,
the characteristic mass, size and velocity dispersion of the central
galaxy are, respectively, $\Mcen\simeq 7.5\times10^{11}\Msun$,
$\rcen\simeq 24\kpc$ and $\sigmacen\simeq 367\kms$.

\section{Numerical experiments}
\label{sec:simu}

%%%%%%%%%%%%%%FIG 1
\begin{figure}
\centerline{\psfig{file=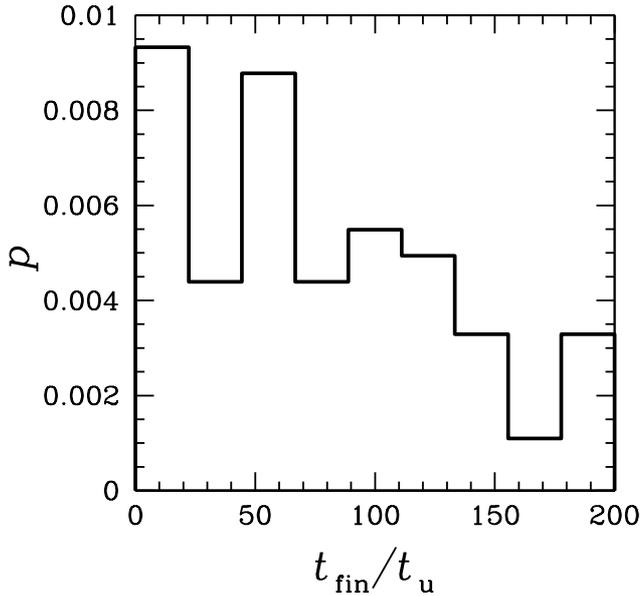,width=\hsize}}
\caption{Probability distribution $p=\d n/\d x$ of the final time
    $\tfin$ defined in Section~\ref{sec:class} ($x=\tfin/\tu$, where
    $\tu$ is the time unit; see Section~\ref{sec:model}) for the
    simulations classified as encounters.}
\label{fig:tfinhisto}
\end{figure}
%%%%%%%%%%%%%%%%%%%%%%%

%%%%%%%%%%%%%%FIG 2
\begin{figure*}
\centerline{
\psfig{file=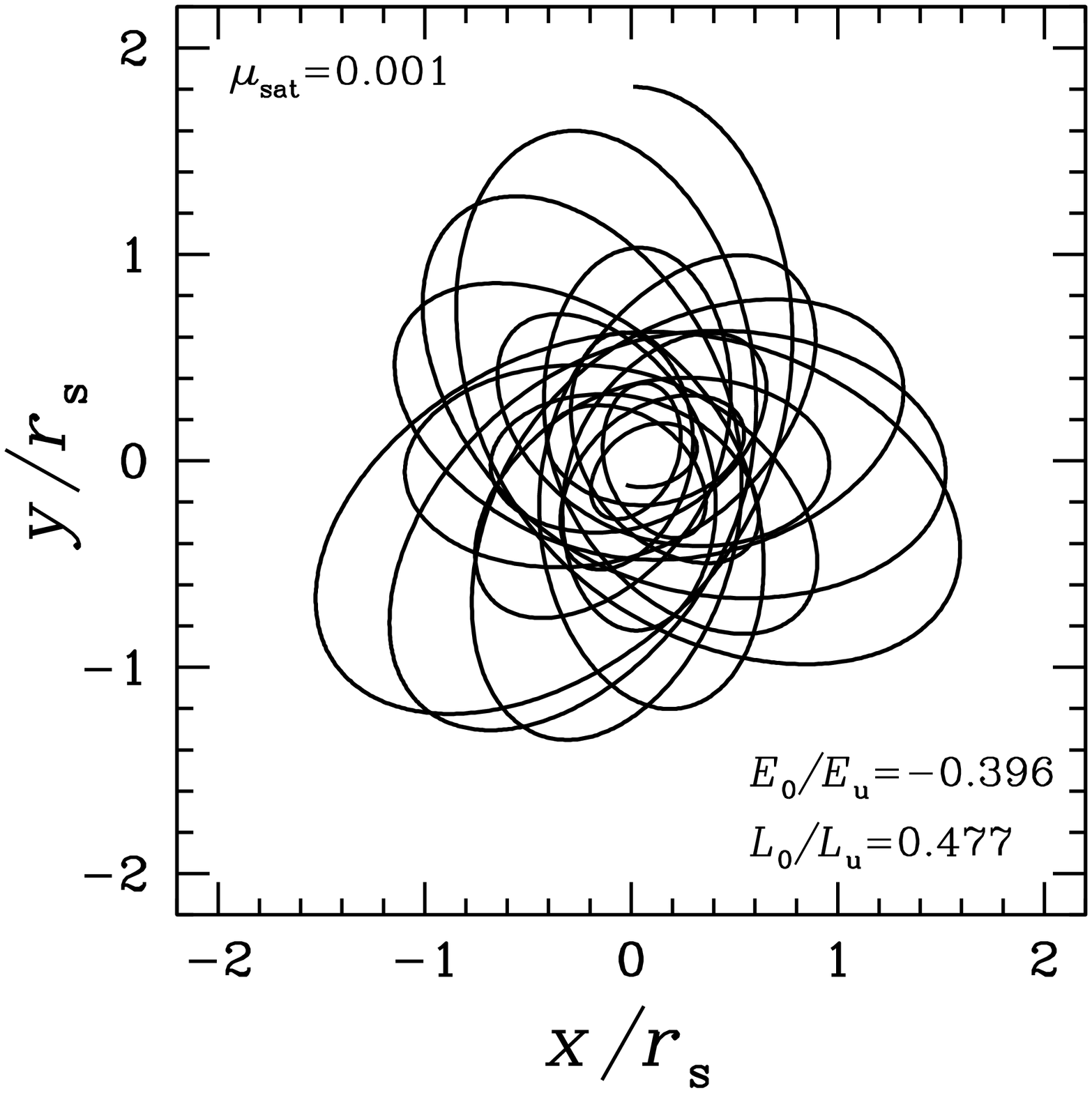,width=0.5\hsize}
\psfig{file=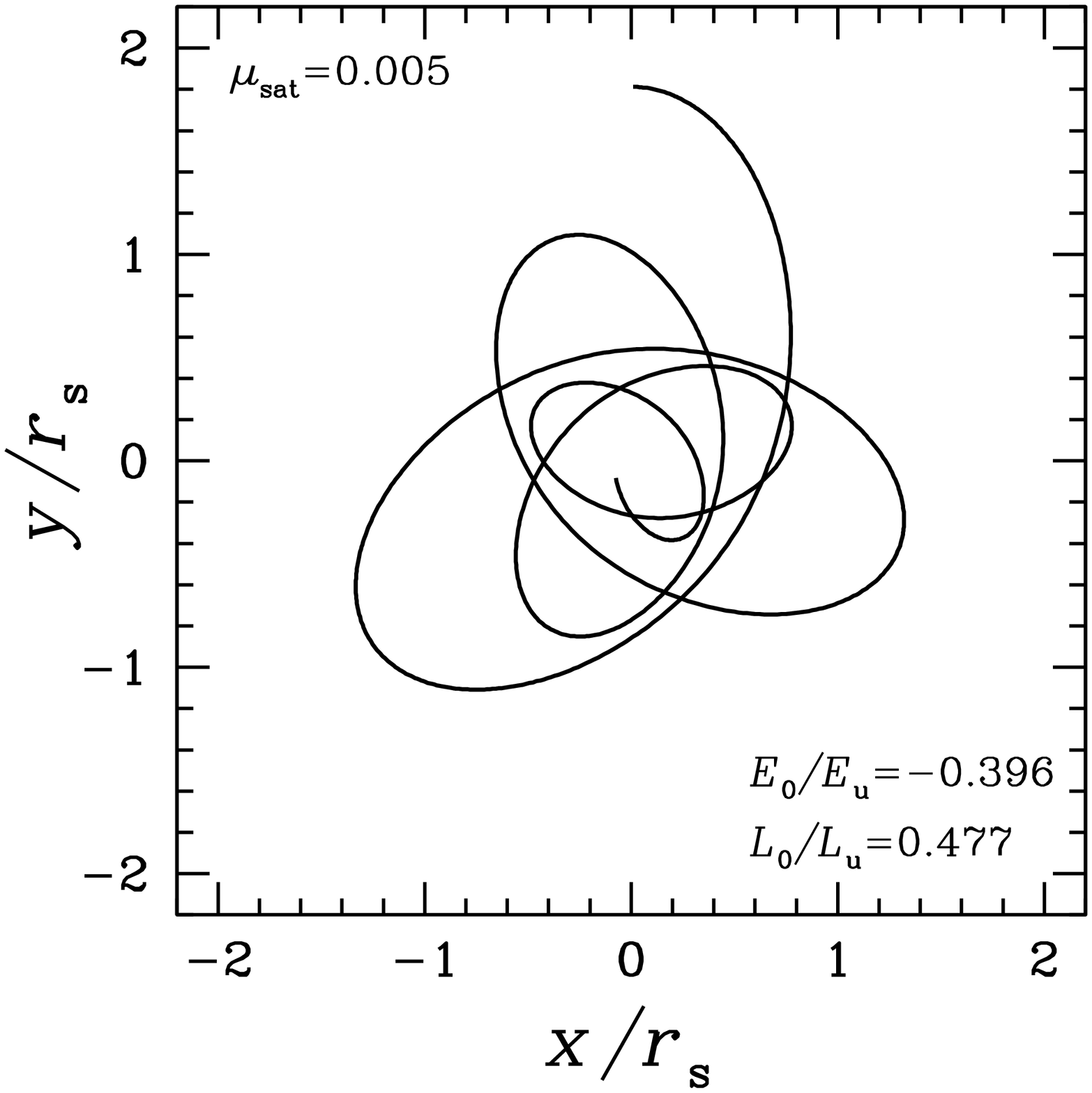,width=0.5\hsize}
}
\centerline{
\psfig{file=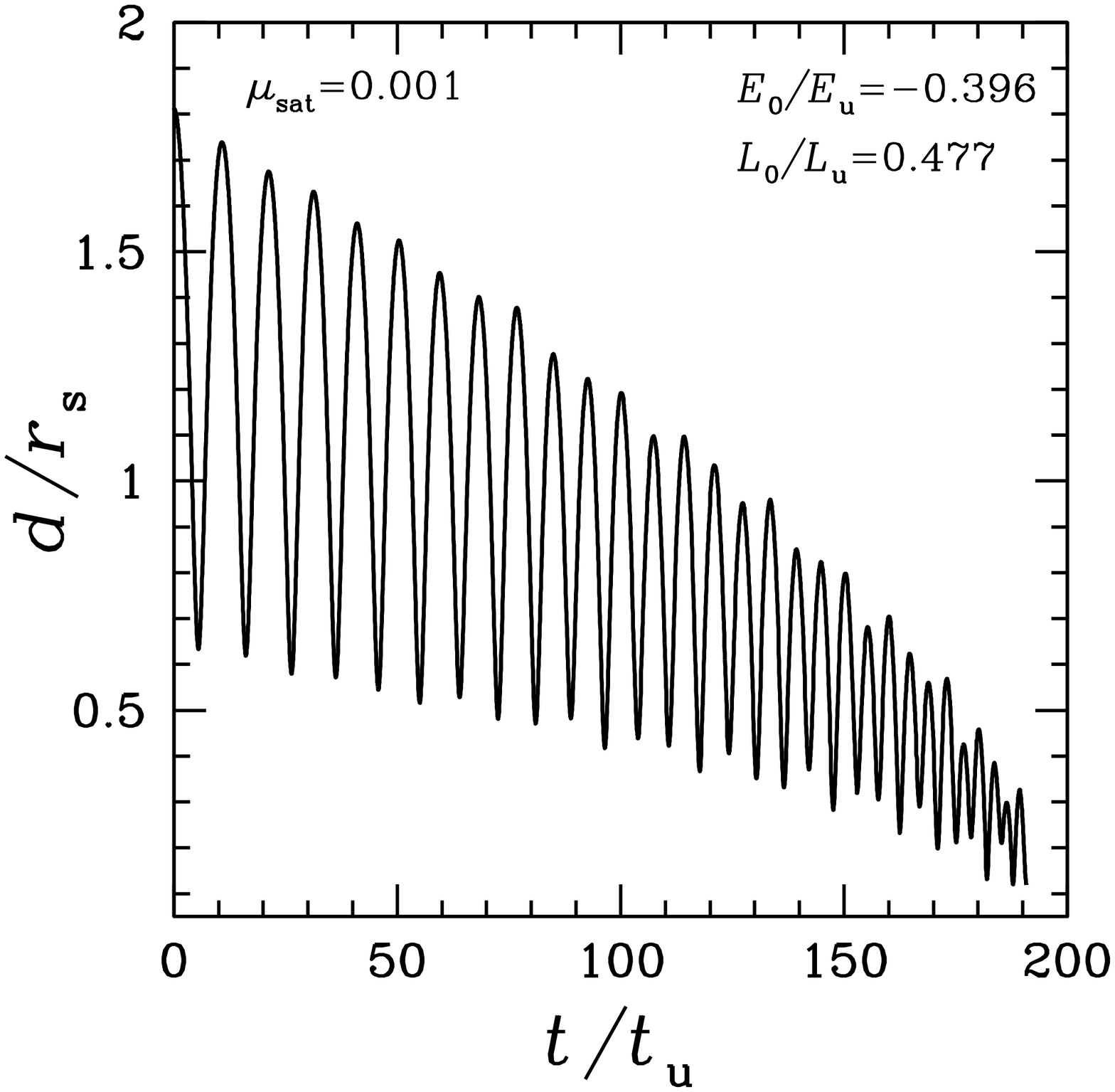,width=0.5\hsize}
\psfig{file=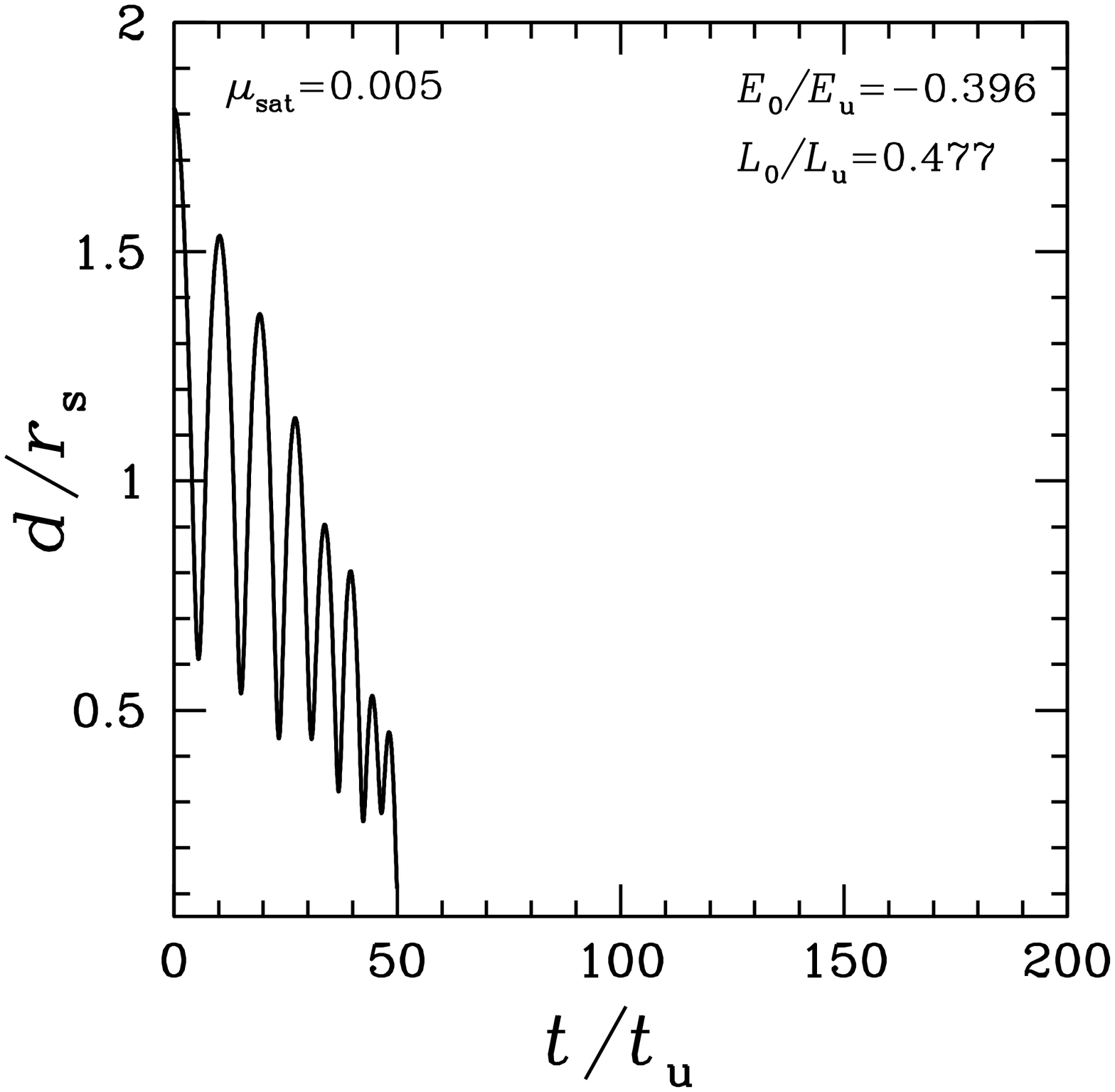,width=0.5\hsize}
}
\caption{{\em Upper panels.} Trajectories of the satellite in the
  $x$-$y$ plane in two representative simulations with the same
  (isotropic; $\ra/\rs=\infty$) DM phase-space distribution and
  initial satellite's orbital parameters ($\Ezero$,$\Lzero$), but
  different satellite's mass: $\musat=0.001$ (left-hand panel) and
  $\musat=0.005$ (right-hand panel).  The trajectories are drawn from
  $t=0$ ($x=0$, $y=d\simeq1.8\rs$) to the final time $\tfin$ when the
  satellite crosses the central sphere of radius $\rcen=0.12\rs$.
  {\em Lower panels.} Distance $d$ of the satellite from the centre of
  the host halo as a function of time for the same two simulations as
  in the upper panels: $\musat=0.001$ (left-hand panel) and
  $\musat=0.005$ (right-hand panel).  $d(t)$ is drawn from $t=0$ to
  the final time $\tfin$, such that $d(\tfin)=\rcen=0.12\rs$. Here
  $\rs$ is the halo scale radius, $\tu$ is the time unit, $\Lu$ is the
  specific angular-momentum unit, and $\Eu$ is the specific energy
  unit (see Section~\ref{sec:model}).}
\label{fig:xydt}
\end{figure*}
%%%%%%%%%%%%%%%%%%%%%%%

%%%%%%%%%%%%%%FIG 3
\begin{figure*}
\centerline{\psfig{file=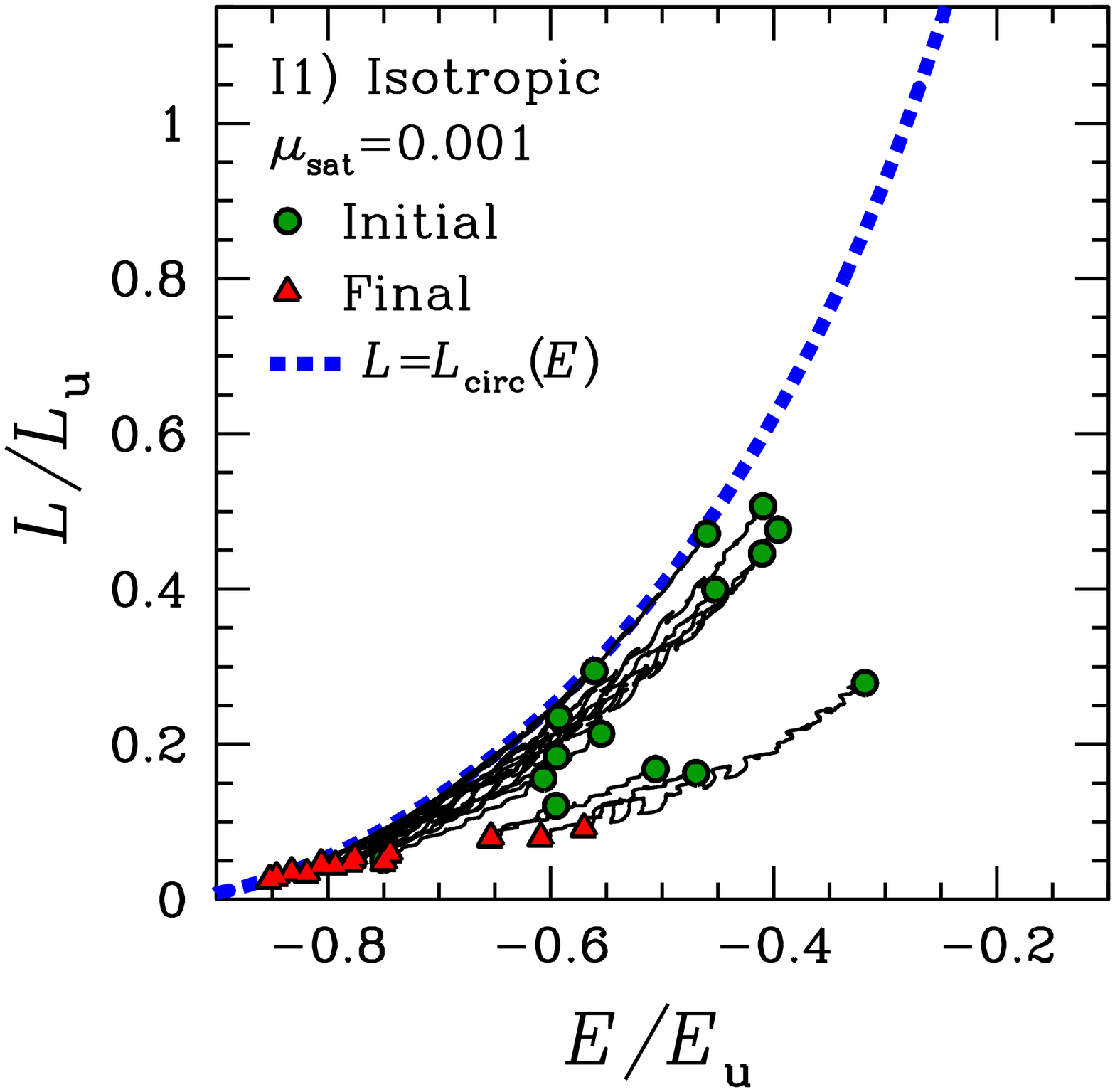,width=0.5\hsize}\psfig{file=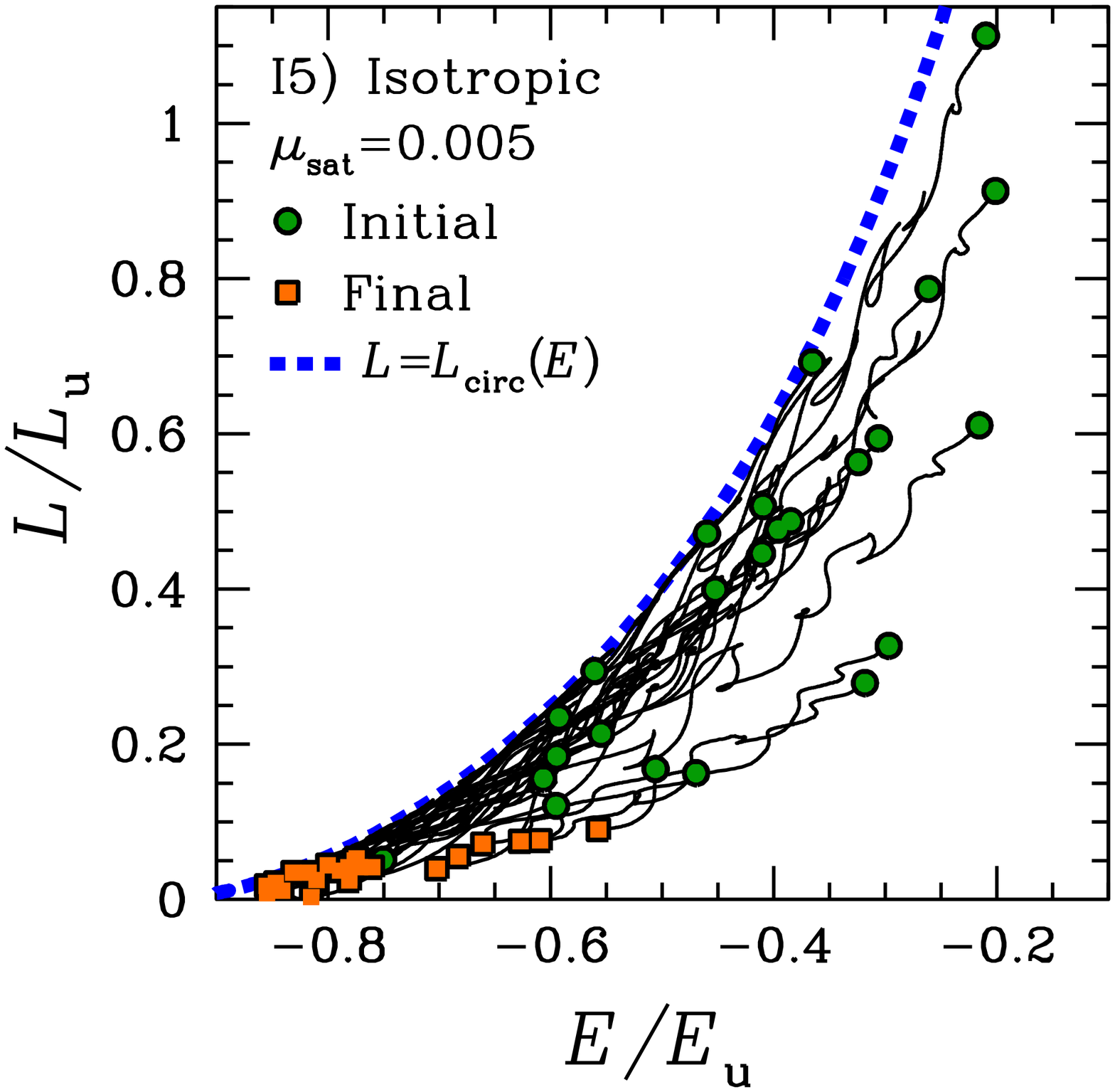,width=0.5\hsize}}
\centerline{\psfig{file=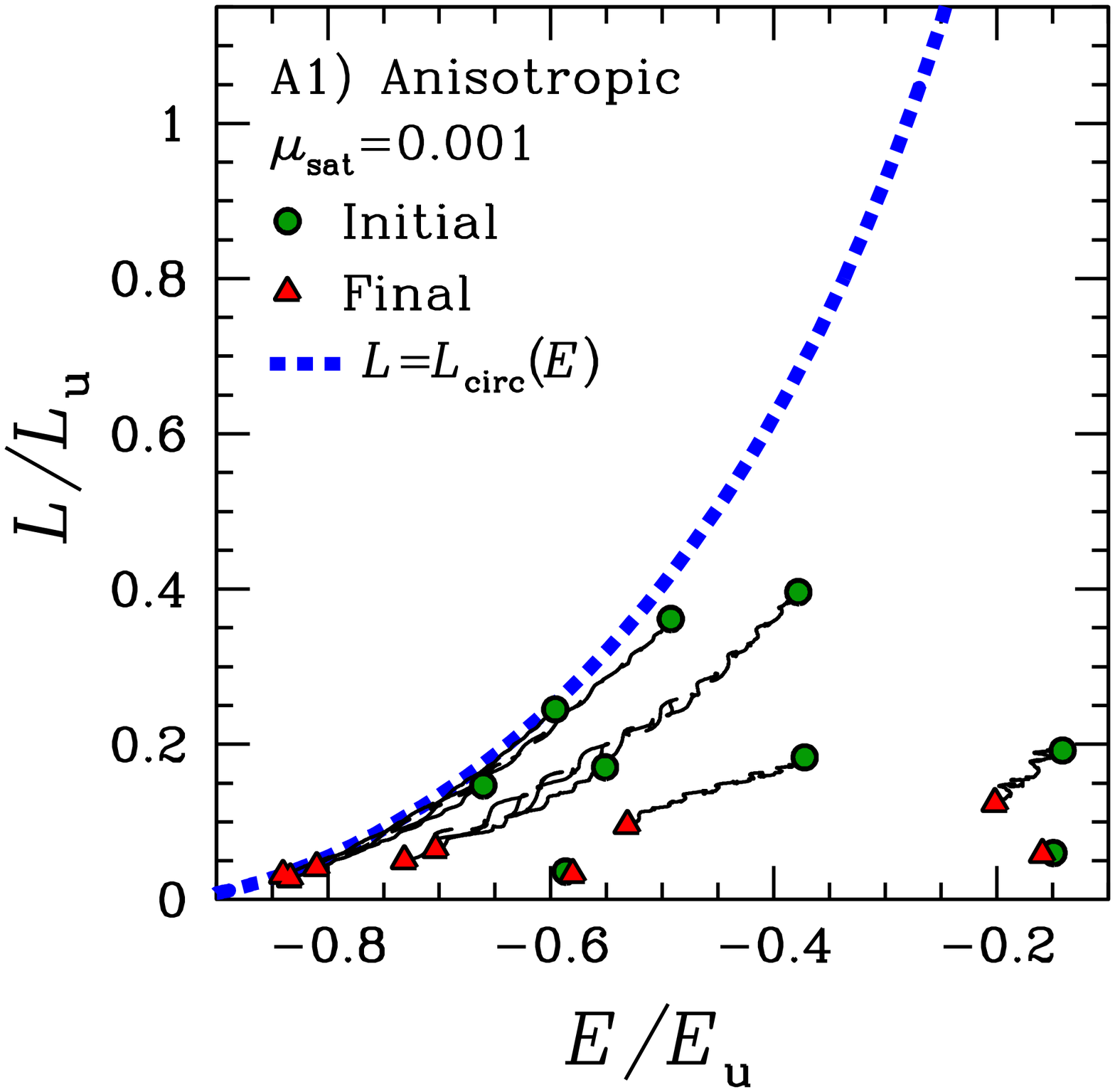,width=0.5\hsize}\psfig{file=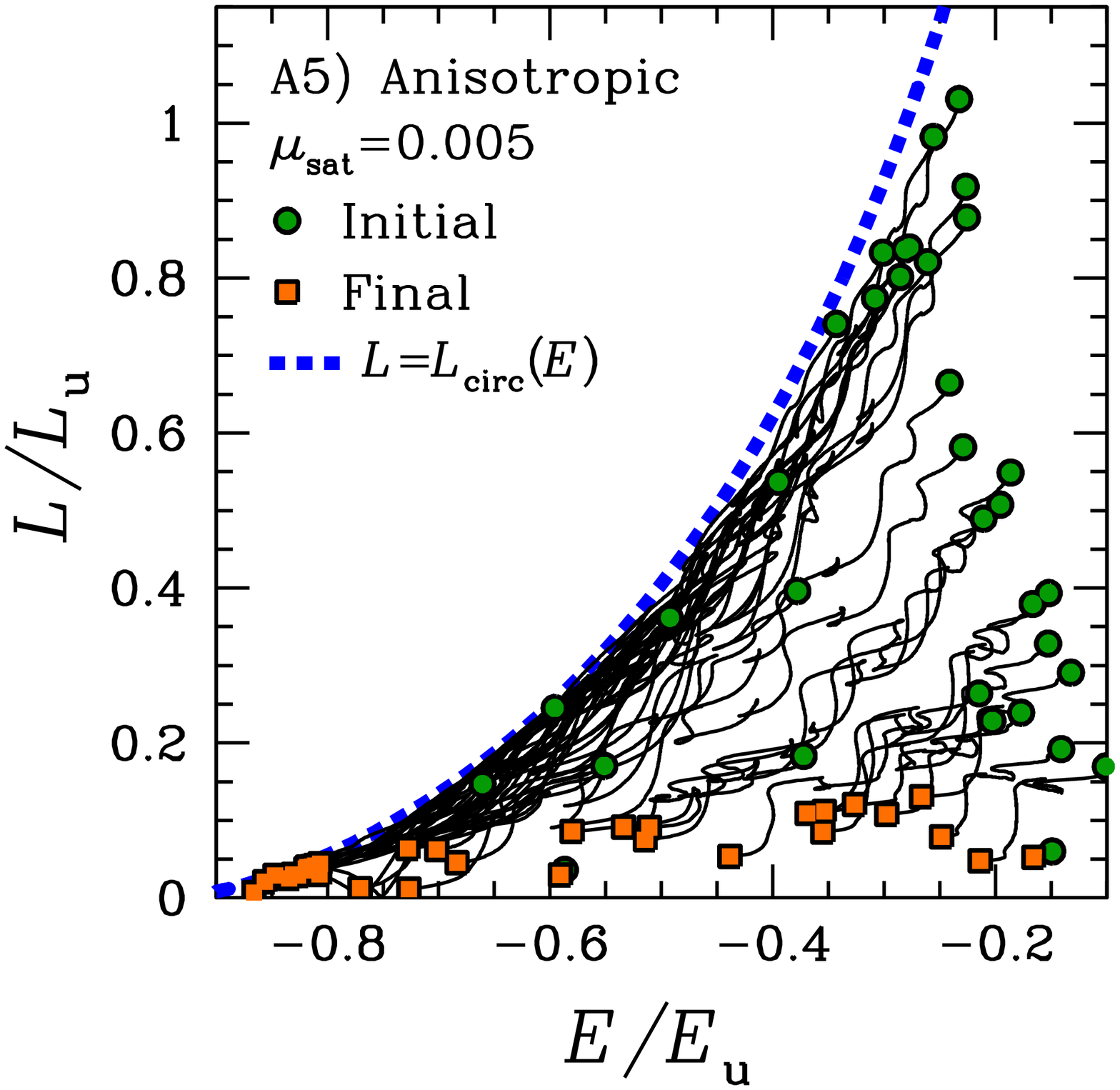,width=0.5\hsize}}
\caption{Initial ($t=0$; circles) and final ($t=\tfin$; triangles and
  squares) specific angular-momentum modulus $L$ versus specific
  energy $E$ of the satellite in the families of models I1, I5, A1 and
  A5 (from top left to bottom right). Only runs that ended up with an
  encounter between the CG and the satellite are plotted. The thin
  solid curves indicate the trajectories of the satellite in the
  $E$-$L$ space. The thick dashed curve indicates the specific
  angular-momentum modulus of a circular orbit with specific energy
  $E$. Here $\Eu$ and $\Lu$ are, respectively, the specific energy and
  angular-momentum units of the model (see Section~\ref{sec:model}).}
\label{fig:el}
\end{figure*}
%%%%%%%%%%%%%%%%%%%%%%%

\subsection{Set-up of the $N$-body simulations}
\label{sec:setup}

In each simulation the initial conditions consist of an equilibrium
spherical DM halo with density distribution~(\ref{eq:rho}), realized
with $N\simeq 10^6$ particles, hosting a massive particle of mass
$\Msat$, representing the satellite.  All DM particles have the same
mass $m=\Mtot/N\simeq10^{-6}\Mtot$. As we follow the evolution of a
single satellite in each simulation, we neglect the effects of
satellite-satellite interactions, which however are not expected to be
very important, as also suggested by studies of the evolution of the
sub-halo population in cosmological halos \citep{van16}.

The $N$-body simulations were run with the parallel collisionless code
\FVFPS \citep[][]{Lon03,Nip03a}.  We adopted the following values of
the code parameters: minimum value of the opening parameter
$\theta_{\rm min}=0.5$, softening parameter for the DM particles
$\varepsilonDM = 0.02 \rs$ and adaptive time-step $\Delta t$ (the same
for all particles, but varying as a function of the maximum mass
density) spanning the range $0.01\lesssim \Delta t/\tu \lesssim 0.1$.

We explore four different families of models obtained by combining two
values of the anisotropy parameter $\ra$ and two values of the
satellite mass $\Msat$. In particular, we explored an isotropic model
($\ra/\rs=\infty$) and a radially anisotropic model ($\ra/\rs=2$):
these two models should bracket the range in orbital anisotropy for
group and cluster-scale DM halos, which are expected to have
moderately radially biased orbits \citep{Asc08,Woj13}.  For each
choice of $\ra$ we explored two values of the satellite mass:
$\musat=0.001$ and $\musat=0.005$, where $\musat\equiv\Msat/\Mtot$ is
the satellite mass in units of the total mass of the host. So, given
our definition of the CG mass (equation~\ref{eq:mcen}), the mass ratio
between the satellite and the central is $\Msat/\Mcen\simeq 0.133$
when $\musat=0.001$ and $\Msat/\Mcen\simeq 0.667$ when $\musat=0.005$.
In the simulations the particle representing the satellite has
softening length $\varepsilonsat$, depending on the satellite mass
$\Msat$: $\varepsilonsat=0.05\rs$ when $\musat=0.001$ and
$\varepsilonsat=0.08\rs$ when $\musat=0.005$. The simulations are run
for a maximum time $\tmax=200\tu$. We verified numerically that the
anisotropic model is stable by running it without the massive
satellite for $200\tu$.

For given initial orbital parameters $\Ezero$ and $\Lzero$, at $t=0$
the satellite is placed at the apocentre of the orbit: taking as
reference a Cartesian system with origin in the halo centre, the
initial coordinates of the satellites are $y=\rapo$, $\vx=\vapo$,
$x=z=\vy=\vz=0$, where $\rapo$ is the apocentric radius and
$\vapo=\sqrt{2[\Ezero-\Phi(\rapo)]}$.  If the satellite had the same
mass as the DM particles ($\Msat=m$) it would orbit in the stationary
potential with constant specific energy $E(t)=\Ezero$ and angular
momentum $L(t)=\Lzero$.  We ran a few test simulations with $\Msat=m$
(and $\varepsilonsat=\varepsilonDM=0.02\rs$) finding that $E$ and $L$
are conserved to within $0.5\%$ and $5\%$, respectively, over the time
span of $200\tu$.

\subsection{Definition of the final  time and classification of the runs}
\label{sec:class}

In the simulations the ratio between satellite mass and DM particle
mass ($\Msat/m=1-5\times 10^3$) is sufficiently large to fully catch
the effect of dynamical friction. The satellite tends to spiral-in
towards the centre of the halo, losing angular momentum ($L$
decreases) and becoming more bound to the host system ($E$
decreases). We monitor the orbit of the satellite and, in particular,
we measure as a function of time the distance $d=\sqrt{x^2+y^2+z^2}$
of the satellite from the halo centre. In all cases, in the initial
conditions $d>\rcen$, where $\rcen$ is the fiducial radius of the CG
(Section~\ref{sec:model}). We define the {\em final time} of the
simulation $\tfin$ as the time when the satellite first crosses the
sphere of radius $\rcen$ centred in the halo centre of mass.  If this
occurs before $\tmax$ we classify the run as an {\em encounter}
(either {\em merger} or {\em fly-by}; see Section~\ref{sec:twobody})
between the satellite and the CG.  When the encounter does not occur
before $\tmax$, the run is excluded from further analysis, because the
satellite is not likely to contribute to accretion onto the CG (we
note that, for instance, $\tmax=200\tu\approx10^{10}\yr$ when
$\rs=100\kpc$ and $\Mtot=10^{14}\Msun$; see equation~\ref{eq:tu}).

\subsection{Properties of the families of simulations}
\label{sec:families}

For $\musat=0.005$ we ran 34 simulations with $\ra/\rs=\infty$ (family
I5) and 34 simulations with $\ra/\rs=2$ (family A5).  In each of these
68 simulations the initial orbit of the satellite is chosen by
extracting randomly the initial energy $\Ezero$ and angular momentum
$\Lzero$ from the distribution function (equation~\ref{eq:df}) of the
host halo. As for the DM particles (see
Section~\ref{sec:setup}), the anisotropy distributions of the
satellites in the two considered models (isotropic and radially
anisotropic with $\ra/\rs=2$) should bracket the anisotropy distribution
expected for satellite galaxies in clusters \citep{Ian12}.

Of the aforementioned 68 simulations, those that ended up with an
encounter between CG and satellite (24 in the isotropic case, 34 in
the anisotropic case) were also rerun with the same initial
conditions, but lower satellite mass $\musat=0.001$ (these families of
simulations are called I1 and A1, respectively\footnote{The ten
  simulations with $\musat=0.005$ that did not end up with an
  encounter were not rerun with $\musat=0.001$, because we know in
  advance that the encounter would not occur for $\musat=0.001$,
  because dynamical friction is less effective for a lower-mass
  satellite.}). Of these 58 simulations with $\musat=0.001$, 24 ended
up with an encounter (9 in the anisotropic case, 15 in the isotropic
case). Thus, altogether we ran 126 simulations, 82 of which ended up
with an encounter (see Table~\ref{tab:par}).  The time $\tfin$
  elapsed before the encounter depends on both the initial orbital
  parameters and the satellite's mass. The distribution of $\tfin/\tu$
  for the 82 simulations classified as encounters, shown in
  Fig.~\ref{fig:tfinhisto}, is quite broad, spanning the range
  $1.3\lesssim \tfin/\tu \lesssim 190$, with mean $\simeq 78$ and
  standard deviation $\simeq 54$.

\begin{table}
%\begin{center}
\caption{Properties of the four families of models (I1, I5, A1,
  A5). $\musat=\Msat/\Mtot$: mass of the satellite in units of the
  total mass of the host system. $\ra/\rs$: ratio between the
  anisotropy radius and the scale radius of the host halo. {\em Runs}:
  number of simulations. {\em Encounters}: number of runs that ended
  up with an encounter (as defined in Section~\ref{sec:class}). {\em
    Mergers}: number of encounters classified as bona-fide
  mergers. {\em Fly-bys}: number of encounters classified as fly-bys
  (Section~\ref{sec:mergerflyby}).\label{tab:par}}
\begin{tabular}{lrrrrrr}
\hline 
Family & $\musat$ & $\ra/\rs$ & Runs & Encounters & Mergers & Fly-bys\\
\hline
I1 & $0.001$ & $\infty$& 24 & 15 & 7 & 8  \\
I5 & $0.005$ & $\infty$ & 34 & 24 & 19 & 5 \\
A1 & $0.001$ & $2$ & 34 & 9  &   3  & 6  \\
A5 & $0.005$ & $2$    &  34 & 34 & 16 & 18 \\
\hline
\end{tabular}
%\end{center}
\end{table}

\section{Results}
\label{sec:results}

\subsection{Morphology of the orbits}

Two examples of the trajectories of simulations in which the satellite
spirals-in down to the CG are shown in the top panels of
Fig.~\ref{fig:xydt}.  For the same two simulations, the bottom panels
of Fig.~\ref{fig:xydt} show the distance $d$ of the satellite from the
halo centre as a function of time. The DM phase-space distribution
(equation~\ref{eq:df} with $\ra/\rs=\infty$) and the initial
conditions of the satellite's orbit are the same in the two
simulations, which differ only for the value of $\musat$. These plots
nicely show that, as well known, dynamical friction is more effective
when $\musat=0.005$ than when $\musat=0.001$. The final time of the
simulation $\tfin$, such that $\d(\tfin)=\rcen$, is $\tfin/\tu\simeq
190$ when $\musat=0.001$ and $\tfin/\tu\simeq 50$ when $\musat=0.005$.

%%%%%%%%%%%%%%FIG 4
\begin{figure}
\centerline{\psfig{file=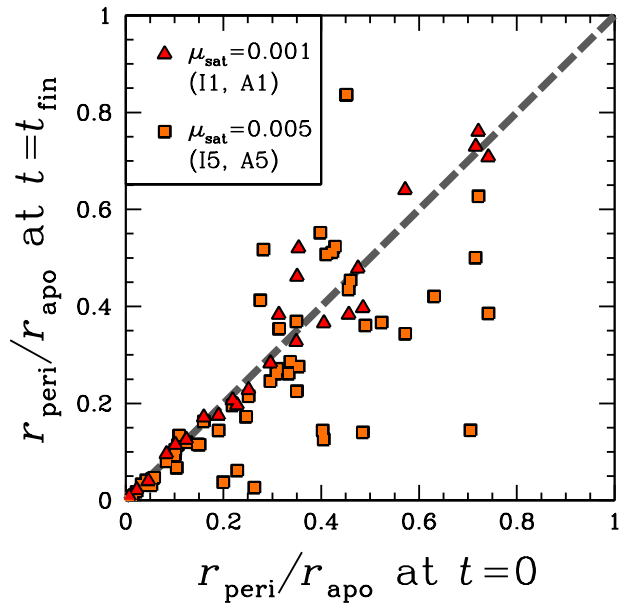,width=\hsize}}
\caption{Final versus initial pericentric-to-apocentric radius ratio
  $\rperi/\rapo$ of the satellite's orbit for simulations that ended
  up with an encounter. The diagonal dashed line indicates no net
  evolution in $\rperi/\rapo$.}
\label{fig:rpra}
\end{figure}
%%%%%%%%%%%%%%%%%%%%%%%

%%%%%%%%%%%%%%FIG 5
\begin{figure}
\centerline{\psfig{file=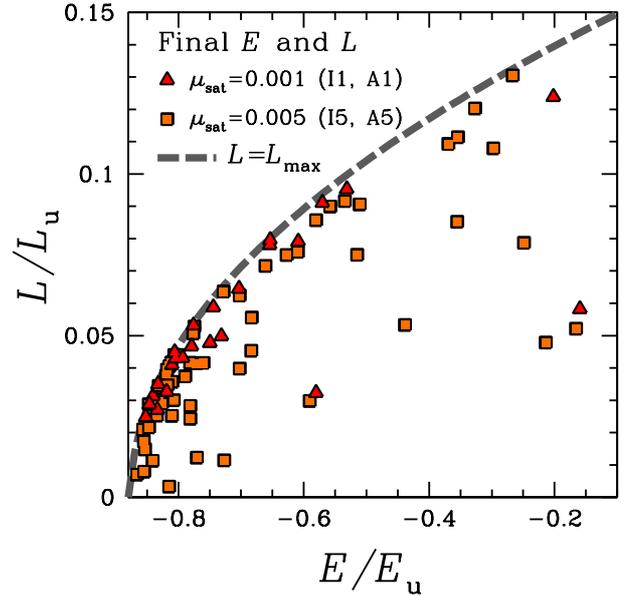,width=\hsize}}
\caption{Final ($t=\tfin$, $d=\rcen$) specific angular-momentum
  modulus $L$ versus specific energy $E$ of the satellite in the same
  units as in Fig.~\ref{fig:el}.  The thick long-dashed curve
  indicates the maximum specific angular-momentum modulus $\Lmax$ of a
  particle with specific energy $E$ located at a distance $d=\rcen$
  from the halo centre (equation~\ref{eq:lmax}).}
\label{fig:elfin}
\end{figure}
%%%%%%%%%%%%%%%%%%%%%%%

\subsection{Evolution of energy and angular momentum}

\subsubsection{Lindblad diagrams}

The effect of dynamical friction can be quantitatively visualised in
the so-called {\em Lindblad diagram} \citep{Lin33}, which is a
representation of the evolution of the orbits in the $E$-$L$ space. It
is useful to define the maximum specific angular-momentum modulus for
an orbit with specific energy $E$, which is the specific
angular-momentum modulus $\Lcirc$ of the circular orbit with the same
specific energy.  The curve $\Lcirc(E)$ delimits the allowed region
$L(E)\leq\Lcirc(E)$ in the Lindblad diagram. Given a spherical
gravitational potential, due to dynamical friction a massive particle
(such as the satellite considered in our model) is expected, in the
Lindblad diagram, to follow tracks in the direction of decreasing
specific energy and angular momentum \citep{Tre75}.  The rate of
evolution along the track increases with the satellite mass (in the
limit $\Msat=m$ the rate becomes zero because $E$ and $L$ are
integrals of motion).  Such tracks are found numerically in our
$N$-body simulations, as apparent from the panels in
Fig.~\ref{fig:el}, showing the initial position, final position and
trajectory of the satellite in the $E$-$L$ plane for the subset of
runs that ended up with an encounter between the satellite and the CG.
We note that the trajectories are smoother for more massive satellites
($\musat=0.005$; right-hand panels in Fig.~\ref{fig:el}) than for less
massive satellites ($\musat=0.001$; left-hand panels in
Fig.~\ref{fig:el}). Figure~\ref{fig:el} also shows that, at the
  final time $\tfin$, the satellites are confined in the bottom area
  of the $E$-$L$ plane: in fact, as we define $\tfin$ as the time when
  the satellite is at a distance $\rcen$ from the centre
  (Section~\ref{sec:class}), for given final $E$ there is an upper
  limit on the final $L$ (see Section~\ref{sec:finel}).

\subsubsection{Orbit circularization?}
\label{sec:circ}

A long-standing question is whether dynamical friction tends to
circularize the orbit of the decelerated object
\citep[e.g.][]{Has03}. Our simulations can provide useful information
to address this question.  In Fig.~\ref{fig:el} circular orbits lie on
the dashed curves: therefore the effect of circularization would be to
bring the orbits closer to these curves. The numerical tracks in the
plots run typically ``parallel'' to these curves, so there is no clear
evidence of circularization.  

A quantitative measure of the deviation from circular orbit is the
pericentric-to-apocentric radius ratio $\rperi/\rapo$, which is unity
for circular orbits and $\ll 1$ for very eccentric orbits. At any
given time $t$ in our simulations we define $\rperi$ and $\rapo$ of
the satellite as the pericentric and apocentric radii of an orbit with
constant specific energy $E$ and angular momentum $L$, where $E$ and
$L$ are those measured at time $t$. The evolution of the
pericentric-to-apocentric radius ratio in our simulations can be
inferred from Fig.~\ref{fig:rpra}, which plots the final ($t=\tfin$)
ratio $\rperi/\rapo$ as a function of the initial ($t=0$) ratio
$\rperi/\rapo$. In the plot the diagonal line corresponds to no net
evolution in $\rperi/\rapo$: if a point lies above (below) this line,
the corresponding final orbit is more circular (eccentric) than the
initial orbit.  Overall, the distribution of points in
Fig.~\ref{fig:rpra} does not suggest that dynamical friction leads to
orbit circularization. In most cases, especially for the lower mass
satellites with $\musat=0.001$, the final and initial $\rperi/\rapo$
are very similar. In some simulations with $\musat=0.005$ the final
$\rperi/\rapo$ is significantly different from the initial value, but
in most of these cases the final orbit is more eccentric than the
initial orbit.

Broadly speaking, our results are in agreement with previous studies,
suggesting that the effect of dynamical friction is not necessarily
orbit circularization.  For instance, \citet{van99} found that
$\rperi/\rapo$ does not evolve significantly for a massive satellite
orbiting in a truncated isothermal halo. The trend observed in
Fig.~\ref{fig:el} is similar to that found by \citet{Are07} in
analogous numerical experiments, but with stellar systems with
different distribution functions. \citet{Are07} conclude that, though
in some cases the effect of dynamical friction is to circularize the
orbit (see also \citealt{Bon87}), this is not a general result, and
that the orbital evolution depends on the properties of the host
stellar system.  We do not find evidence of significantly different
evolution of $\rperi/\rapo$ in simulations with anisotropic (A1 and
A5) and isotropic (I1 and I5) distribution functions, as instead found
by \citet{Cas87} and \citet{Tsu00}, who however studied different
families of stellar systems. Moreover, it must be stressed that,
unlike \citet{Cas87} and \citet{Tsu00}, in our experiments we are not
comparing the evolution of orbits with the same initial orbital
parameters in isotropic and radially anisotropic host systems, because
in our simulations the initial orbital parameters of the satellite are
extracted from the same distribution function as that of the host halo
(see Section~\ref{sec:model}). As suggested by \citet{Sta91}, the host
system tries to make the satellites conform to its orbital structure,
consistent with the fact that in our simulations there is not a net
trend towards orbit circularization, in either the isotropic or the
anisotropic case.

In summary, the results of our simulations add further evidence that
dynamical friction does not necessarily lead to orbit circularization:
the effect on the orbit of the decelerated object depends not only on
the host's distribution function, but also, for given distribution
function, on the initial orbital parameters of the satellite.

%%%%%%%%%%%%%%FIG 6
\begin{figure}
\centerline{\psfig{file=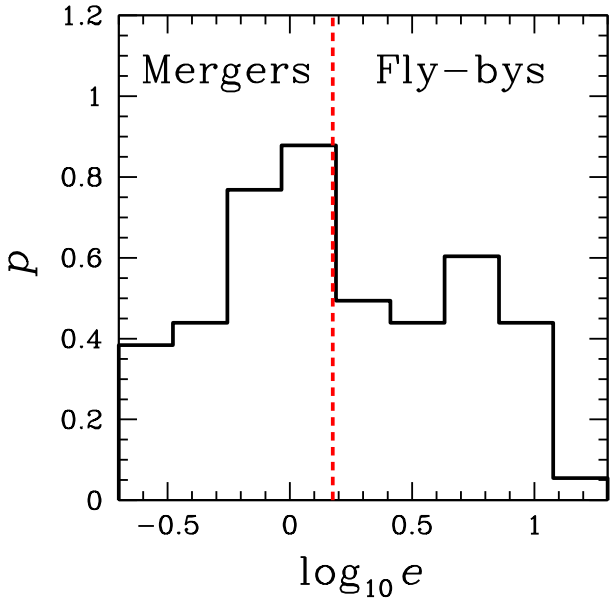,width=\hsize}}
\caption{Probability distribution $p=\d n /\d \log_{10} e$ of the
  final ($t=\tfin$, $d=\rcen$) eccentricity $e$ of central-satellite
  encounters. The vertical dashed line ($e=1.5$) separates {\it
    bona-fide} mergers ($e<1.5$) from fly-bys ($e>1.5$).}
\label{fig:ecchisto}
\end{figure}
%%%%%%%%%%%%%%%%%%%%%%%

%%%%%%%%%%%%%%FIG 7
\begin{figure*}
\centerline{\psfig{file=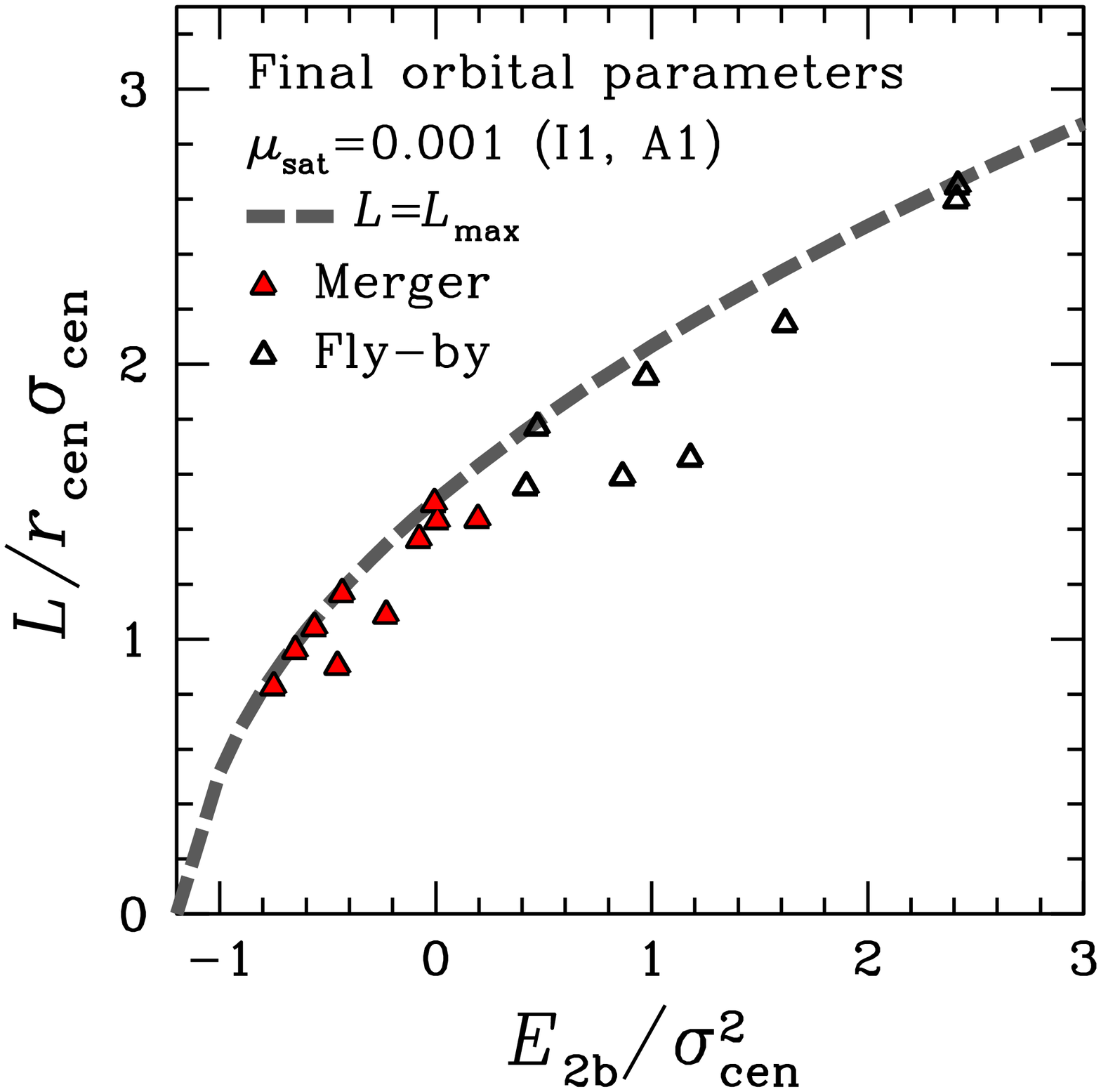,width=0.5\hsize}\psfig{file=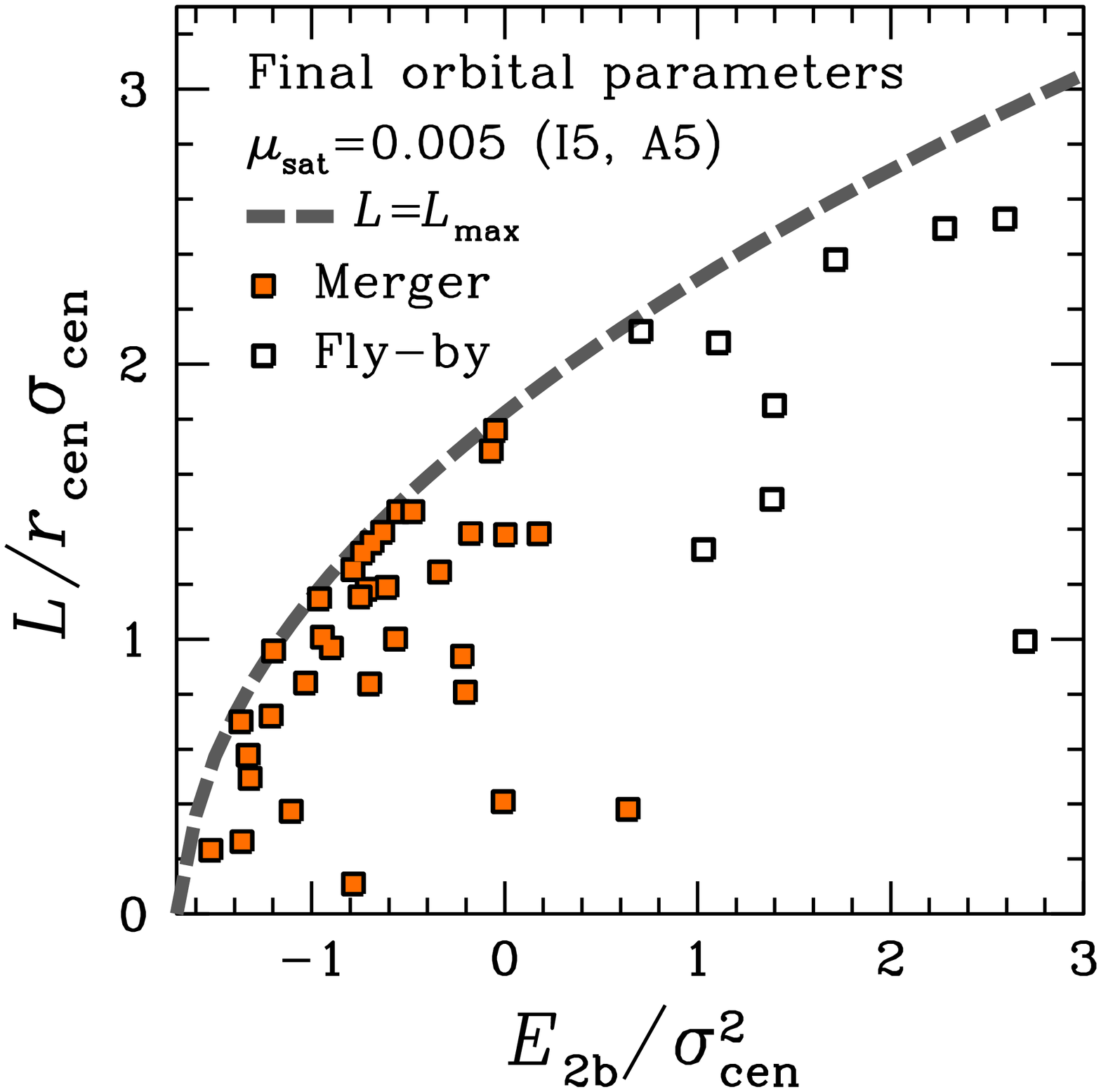,width=0.5\hsize}}
\caption{Final ($t=\tfin$, $d=\rcen$) specific energy (in the two-body
  approximation) $\Etwob$ and angular-momentum modulus $L$ of the
  central-satellite encounter for $\musat=0.001$ (left-hand panel) and
  $\musat=0.005$ (right-hand panel). Solid symbols represent {\it
    bona-fide} mergers, while empty symbols represent fly-bys.  The
  thick long-dashed curve indicates the maximum specific
  angular-momentum modulus of an encounter with specific energy
  $\Etwob$ and satellite-central separation $d=\rcen$
  (equation~\ref{eq:lmax}).  Here $\sigmacen$ and $\rcen$ are,
  respectively, the characteristic velocity dispersion and size of the
  CG (see Section~\ref{sec:model}).}
\label{fig:e2blfin}
\end{figure*}
%%%%%%%%%%%%%%%%%%%%%%%

\subsubsection{Final specific energy and angular momentum}
\label{sec:finel}

The final ($t=\tfin$) positions of the satellites in the $E$-$L$
diagram are shown in Fig.~\ref{fig:elfin}. We recall that, by
definition, these final points are such that the distance of the
satellite from the centre is $d=\rcen$.  At fixed position and energy
of the satellite, the maximum allowed specific angular momentum is in
general smaller than $\Lcirc$. For given distance from the centre of
the host halo $d$ and given specific energy $E$, the maximum value of
$L$ is
\begin{equation}
\Lmax=d\sqrt{2[E-\Phi(d)]}=dv,
\label{eq:lmax}
\end{equation}
where $v$ is the speed of the satellite when it is at a distance $d$
from the centre.  Clearly, $\Lmax\leq\Lcirc(E)$ and $\Lmax$ is such
that $L/\Lmax=\vtan/v$, where $\vtan$ is the tangential velocity.

From Fig.~\ref{fig:elfin} it is apparent that the allowed region in
the final $E$-$L$ space is not uniformly populated, but the points
tend to cluster close to the maximum value $\Lmax$. This suggests that
when the satellite encounters the CG it preferentially has almost the
maximum angular momentum allowed for its energy.  We have seen in
Section~\ref{sec:circ} that dynamical friction does not always tend to
circularize the orbit and that, in any case, the effect of dynamical
friction on the orbital eccentricity is typically small (see
Fig.~\ref{fig:rpra}). The fact that the final $L$ is often close to
$\Lmax$ is not due to orbit circularization, but to the fact that,
given the small size of the CG compared to the host system, it is more
likely that the encounter happens on an orbit ``grazing'' the CG than
on a very radial orbit penetrating the CG close to the centre.  This
is due to the fact that only a small fraction of the system's orbits
go directly through the host's core. The majority of orbits are less
eccentric and shrink as a consequence of dynamical friction, leading
to almost tangential encounters between the satellite and the CG.

%%%%%%%%%%%%%%FIG 8
\begin{figure*}
\centerline{\psfig{file=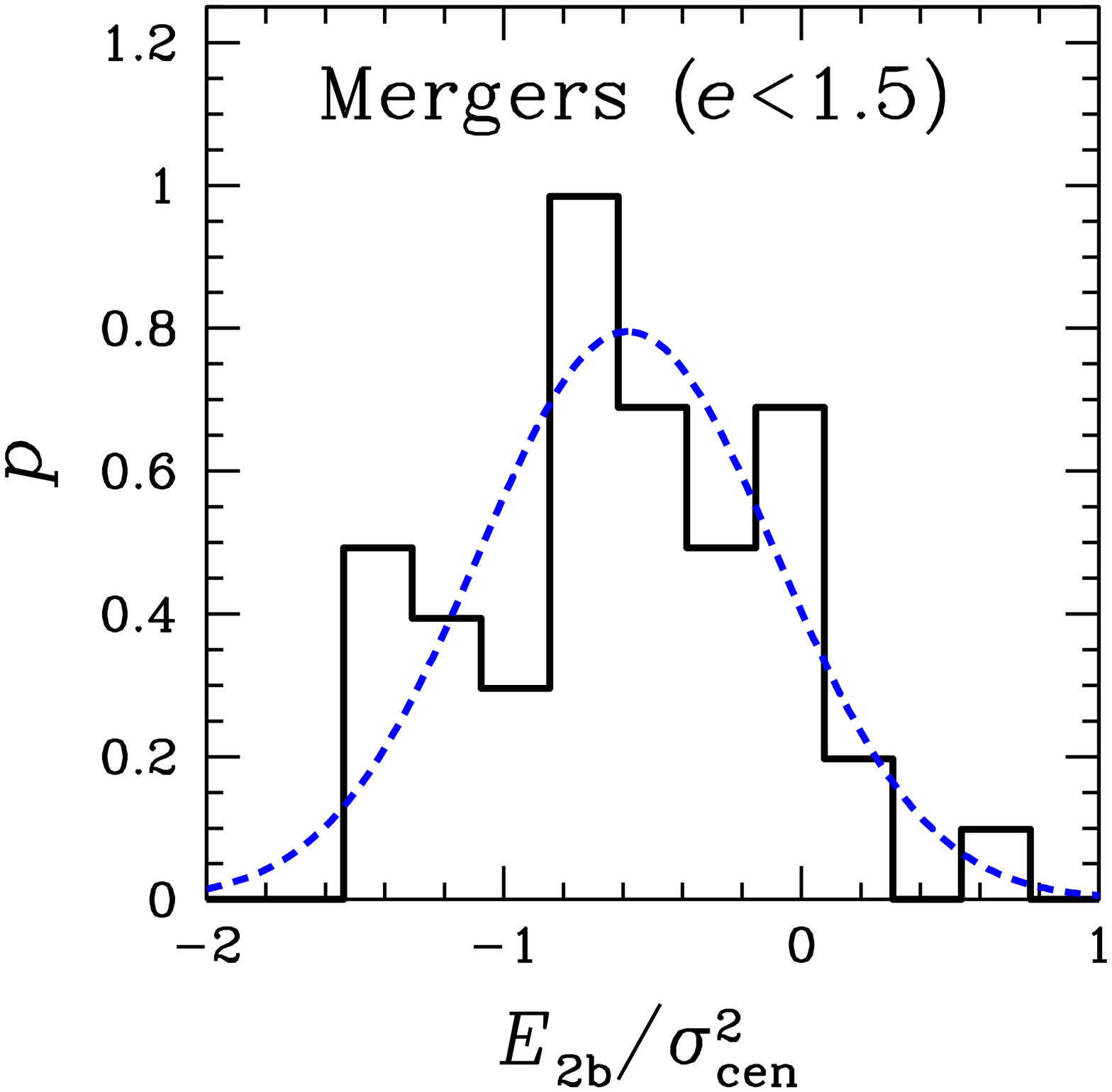,width=0.5\hsize}\psfig{file=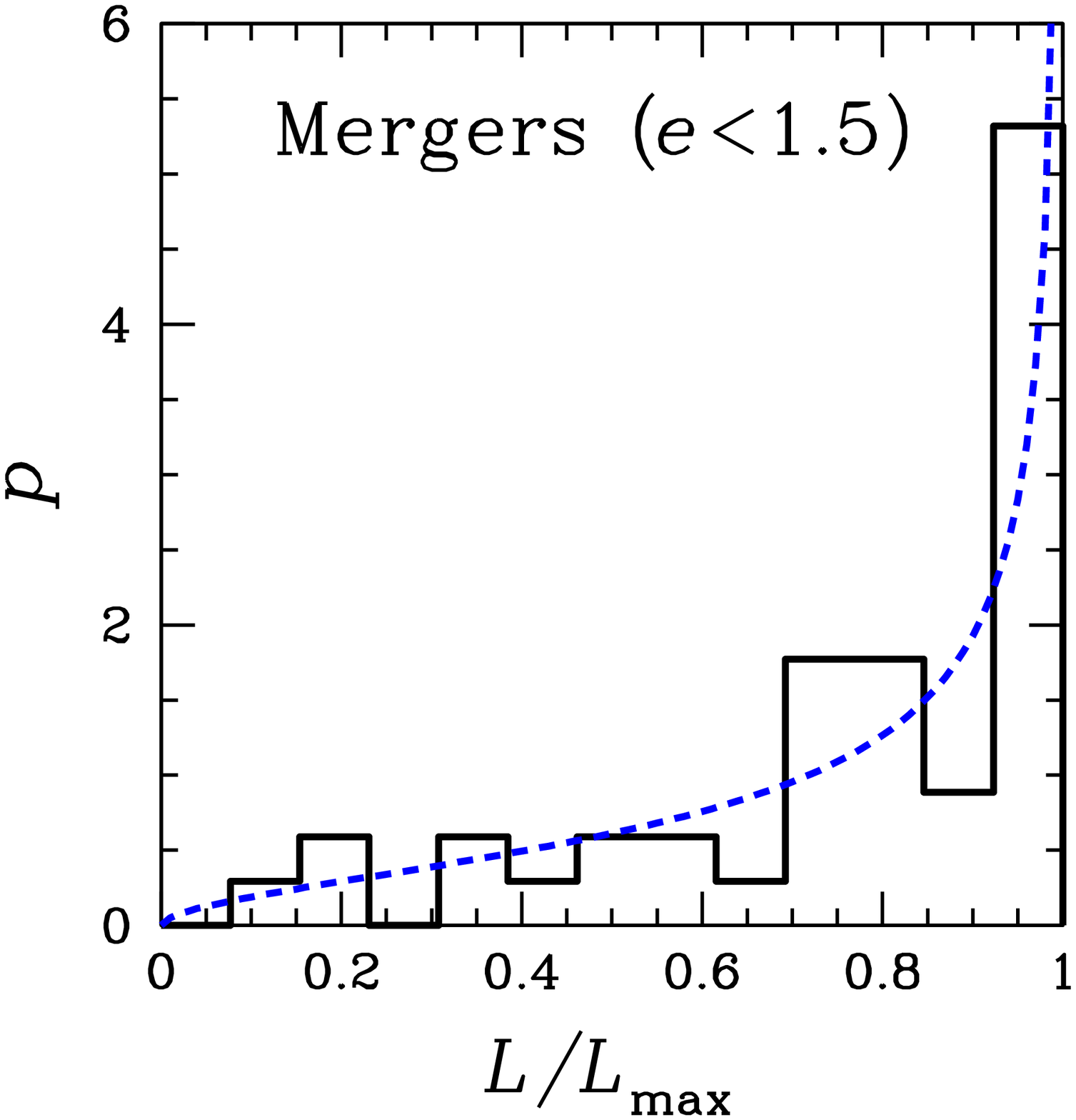,width=0.5\hsize}}
\caption{Probability distribution $p=\d n/\d x$ of the final specific
  orbital energy, computed in the two-body approximation
  ($x=\Etwob/\sigmacen^2$; left-hand panel, solid histogram), and
  angular-momentum modulus ($x=L/\Lmax$; right-hand panel, solid
  histogram), where $\sigmacen$ is the characteristic velocity
  dispersion of the CG and $\Lmax$ is defined by
  equation~(\ref{eq:lmax}). In each panel the dashed curve represents
  the analytic best fit of the distribution (see text).}
\label{fig:e2blhisto}
\end{figure*}
%%%%%%%%%%%%%%%%%%%%%%%

\begin{table*}
%\begin{center}
\caption{Mean ($\mu$) and standard deviation ($\sigma$) of the
  distributions of the orbital parameters ($\eta$, $\rperitwob/\rcen$,
  $v/\vcirc$ and $|\vr|/v$) for the central-satellite mergers of our
  simulations (CGs) and for accretion of cosmological satellites (DM
  halos). The data for cosmological satellites are taken from
  \citet{Wet11} and \citet[][]{Jia15}. \label{tab:dist}}
\begin{tabular}{lcccccccc}
\hline 
 & $\mu(\eta)$ & $\sigma(\eta)$ & $\mu(\rperitwob/\rcen)$ & $\sigma(\rperitwob/\rcen)$  & $\mu(v/\vcirc)$ & $\sigma(v/\vcirc)$  & $\mu(|\vr|/v)$ & $\sigma(|\vr|/v)$  \\
\hline
CGs      &0.64  &0.28  &0.50 &0.32  &1.39  &0.35 &0.53 &0.27   \\
DM Halos &0.51  &0.22  &0.24 &0.22  &1.24  &0.15 &0.79 &0.19   \\
\hline
\end{tabular}
%\end{center}
\end{table*}

\subsection{Central-satellite orbital parameters in the two-body approximation}
\label{sec:twobody}

When an encounter between the central and the satellite occurs it is
useful to describe it in terms of the orbital parameters calculated in
the point-mass two-body approximation. This formalism, though not
rigorous for extended objects, is often used in the study and
classification of mergers of galaxies \citep[e.g.][]{Boy06,Nip09} and
DM halos \citep[e.g.][]{Kho06,Pos14}.

From the point of view of the central-satellite galaxy orbit, we
define the two-body approximation orbital energy per unit mass
\begin{equation}
\Etwob=\frac{1}{2}v^2-\frac{G\Mtwob}{d},
\end{equation}
where $\Mtwob\equiv \Mcen+\Msat$, $d$ is the relative distance and $v$
is the relative speed. As the CG is at rest in the centre of the halo,
$d$ and $v$ are, respectively, the distance and the speed of the
satellite relative to the halo centre. Therefore, the specific orbital
angular momentum per unit mass is just the specific angular momentum
of the satellite, with modulus $L=d\vtan$.

In the two-body approximation the orbit can be characterized also by
different combinations of the two parameters $\Etwob$ and $L$. In
particular, it is useful to define the eccentricity
\begin{equation}
e = \sqrt{1+\frac{2\Etwob L^2}{G^2\Mtwob^2}}
\end{equation}
and the pericentric radius
\begin{equation}
\rperitwob =\frac{G\Mtwob}{2\Etwob}(e-1).
\end{equation}
For bound orbits ($\Etwob<0$, i.e. $e<1$) we define also the circularity
\begin{equation}
\eta = \sqrt{1-e^2}.
\end{equation}

\subsubsection{Bona-fide mergers and fly-bys}
\label{sec:mergerflyby}

Not in all the runs that ended up with an encounter we expect to have
a rapid merger between the central and the satellite galaxies. Rapid
mergers occur when the orbits are bound $\Etwob<0$, but also for
unbound orbits ($\Etwob\geq 0$), provided the orbital angular-momentum
modulus $L$ is sufficiently low \citep[see section 7.4 of][]{Bin87}.
For this reason, a convenient parameter that can be used to identify
rapid mergers is the eccentricity $e$, which, for $\Etwob>0$, is an
increasing function of both $\Etwob$ and $L$. We take as fiducial
discriminating value of eccentricity $e=1.5$ and classify an encounter
as a {\it bona-fide merger} when $e<1.5$ and as a {\em fly-by} when
$e>1.5$.  The distribution of $e$ for our runs that ended up with an
encounter is shown in Fig.~\ref{fig:ecchisto}. The distribution of the
mergers is characterized by a peak around $e=1$. A substantial
fraction of the fly-bys has very high eccentricity (up to $e\gtrsim
10$): in these cases the satellite, though bound to the deep potential
well of the host DM halo, appears as highly unbound from the point of
view of the central-satellite interaction.

Figure~\ref{fig:e2blfin} plots the final position of the satellites in
the $\Etwob$-$L$ plane.  Similar to Fig.~\ref{fig:elfin}, also in this
case the separation $d$ between central and satellite is fixed
$d=\rcen$, so the allowed region of the $\Etwob$-$L$ plane is
$L\leq\Lmax$, where $\Lmax=d v$ (equation~\ref{eq:lmax}), where $v$ is
the satellite's speed when $d=\rcen$.  Similar to
Fig.~\ref{fig:elfin}, also in Fig.~\ref{fig:e2blfin} the points (in
particular those representing mergers) tend to cluster close to the
curve $\Lmax(\Etwob)$, indicating high orbital angular momentum for
the accreted satellites. The orbital energy for bona-fide mergers is
found in the range $-1.5\lesssim \Etwob/\sigmacen^2\lesssim 1$, where
$\sigmacen$ is the characteristic velocity dispersion of the CG
(Section~\ref{sec:model}).

%%%%%%%%%%%%%%FIG 9
\begin{figure*}
\centerline{\psfig{file=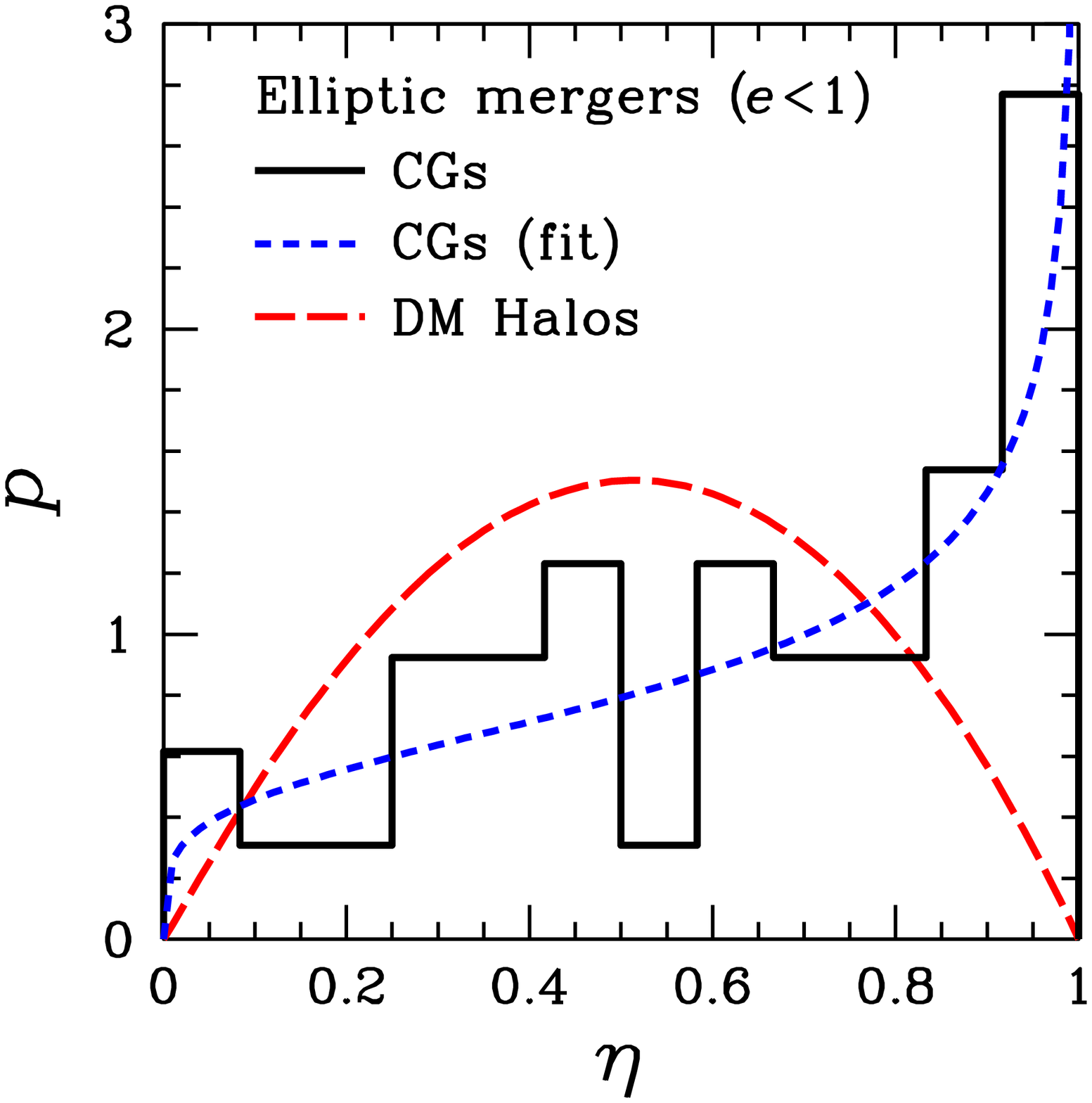,width=0.5\hsize}\psfig{file=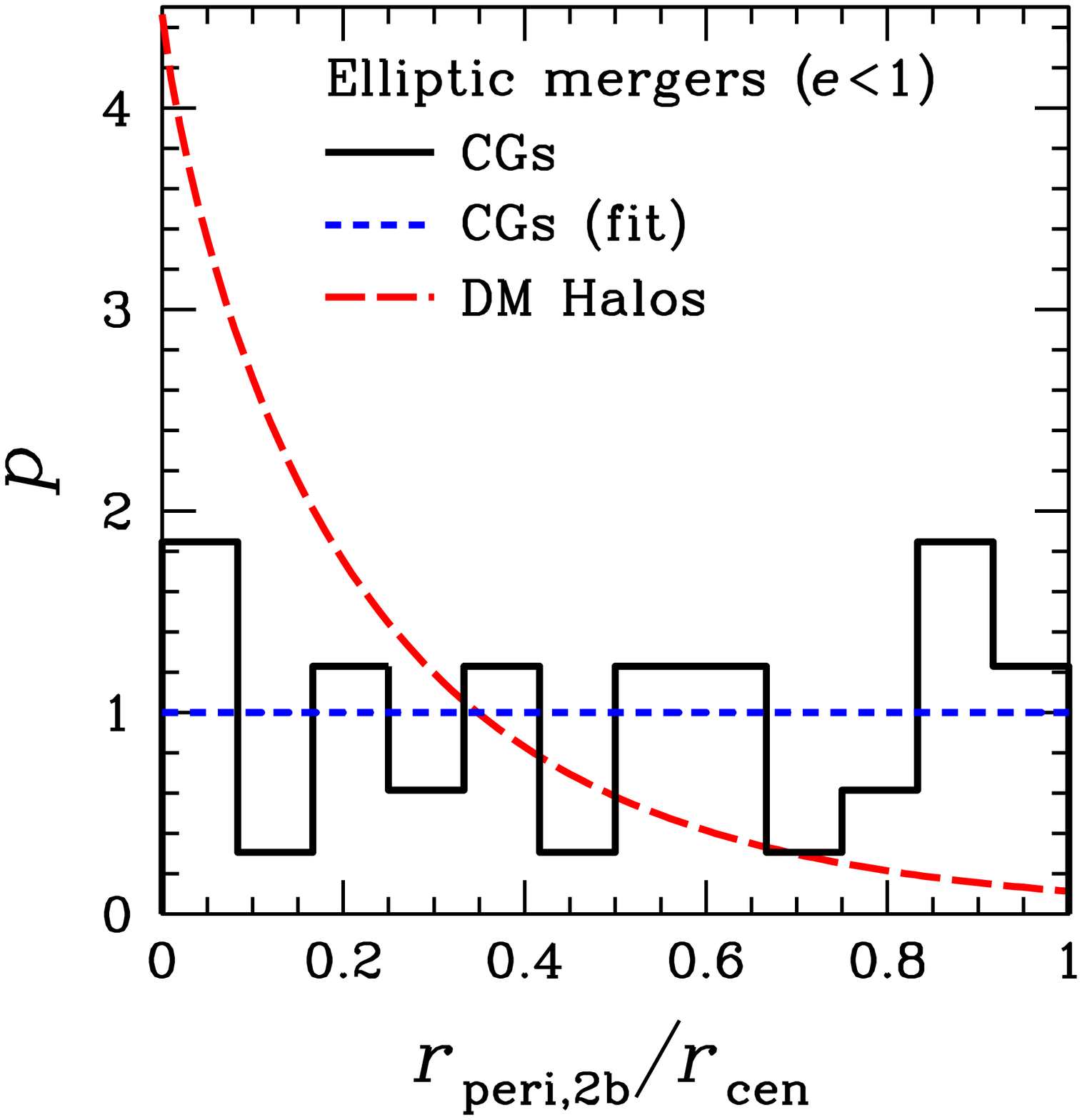,width=0.5\hsize}}
\caption{Probability distribution $p=\d n/\d x$ of the final orbital
  circularity ($x=\eta$; left-hand panel, solid histogram) and
  pericentric radius ($x=\rperitwob/\rcen$; right-hand panel, solid
  histogram) for elliptic mergers in the simulations of the present
  work. In each panel the short-dashed curve represents the analytic
  best fit of the distribution, while the long-dashed curve represents
  the distribution found for DM halos by \citet[][specifically for
    $z=0$ and host-halo mass $10^{13.5}\Msun$; see section 6 in that
    paper]{Wet11}.  In the right-hand panel the results of
  \citet{Wet11} are rescaled by identifying the virial radius with
  $\rcen$.}
\label{fig:etarperihisto}
\end{figure*}
%%%%%%%%%%%%%%%%%%%%%%%

%%%%%%%%%%%%%%FIG 10
\begin{figure*}
\centerline{\psfig{file=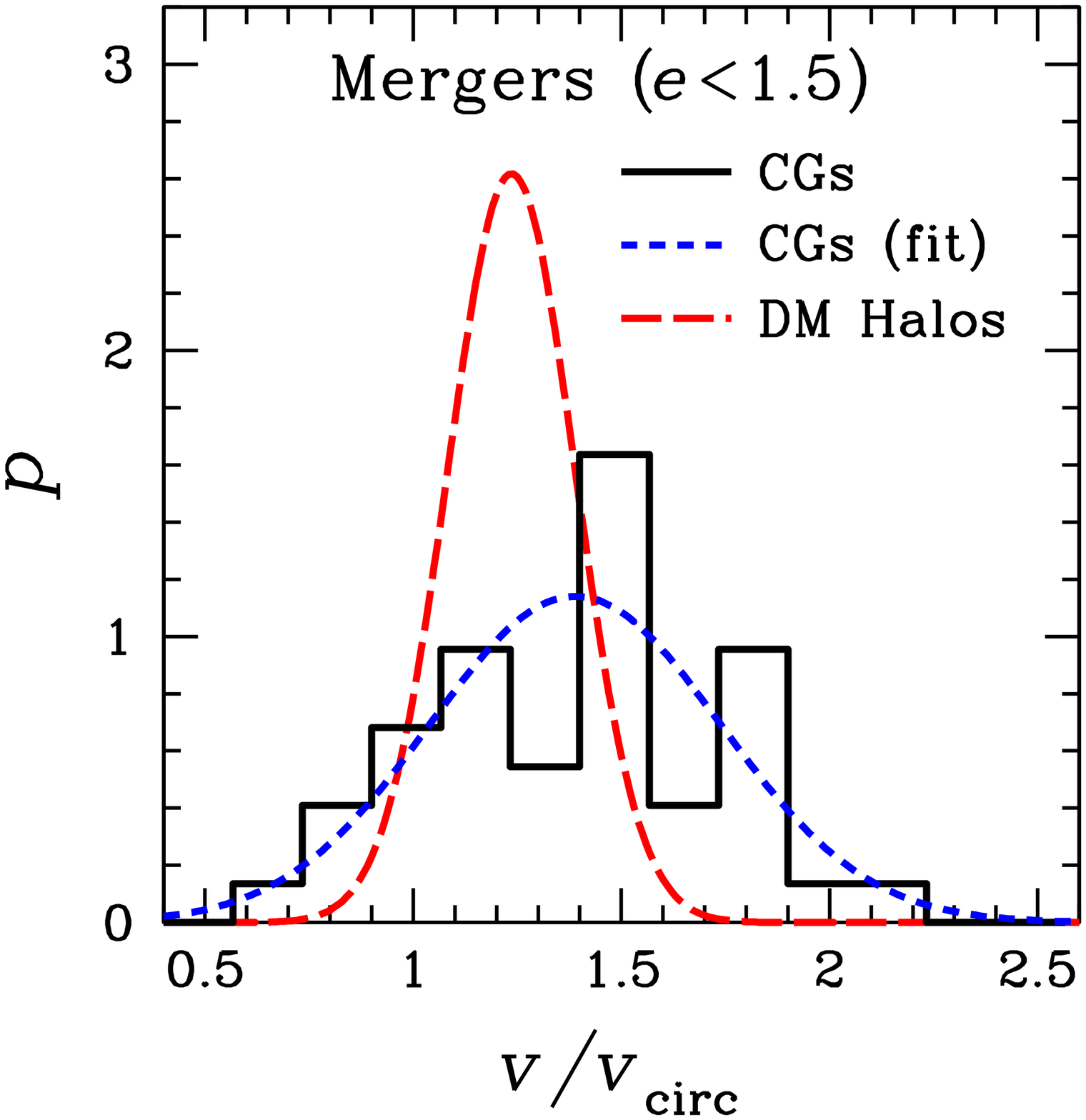,width=0.5\hsize}\psfig{file=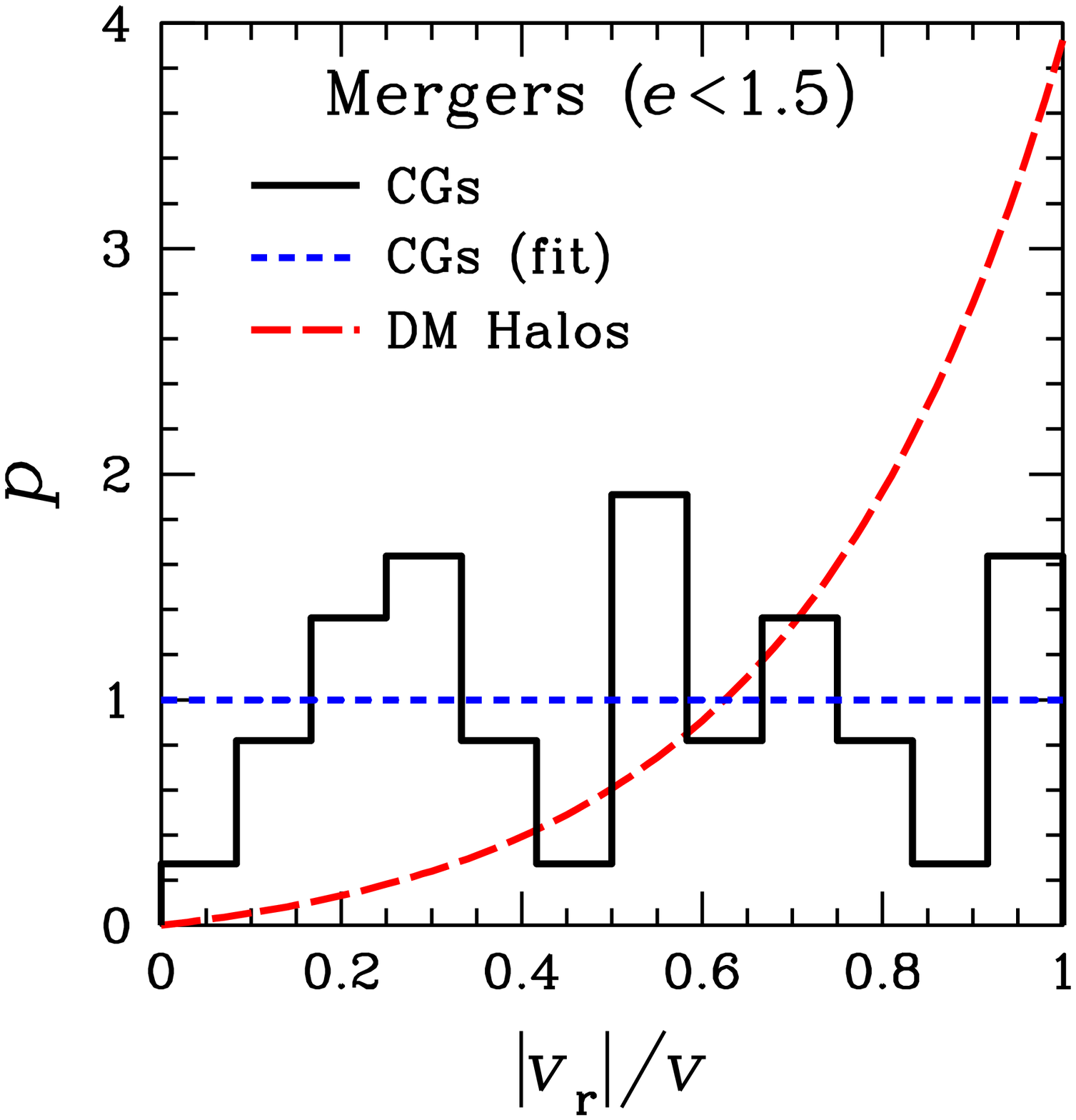,width=0.5\hsize}}
\caption{Probability distribution $p=\d n/\d x$ of the final speed
  ($x=v/\vcirc$; left-hand panel, solid histogram) and radial velocity
  component ($x=|\vr|/v$; right-hand panel, solid histogram) for
  mergers in the simulations of the present work. In each panel the
  short-dashed curve represents the analytic best fit of the
  distribution, while the long-dashed curve represents the
  distribution found for DM halos by \citet[][specifically host-halo
    mass $10^{14}\Msun$ and satellite-to-host mass ratio 0.005-0.05;
    see section 3.4 in that paper]{Jia15}. For the CGs studied in this
  paper we take as reference radius $\rcen$, so $\vcirc=\sigmacen$.
  For the cosmological halos of \citet{Jia15} $\vcirc$ is the circular
  velocity at the virial radius, at which $v$ and $\vr$ are measured.}
\label{fig:vvr}
\end{figure*}
%%%%%%%%%%%%%%%%%%%%%%%

\subsection{Distributions of the orbital parameters of the bona-fide mergers}

%{\bf [ALL KS]}

We focus here on the bona-fide mergers as defined and selected in
Section~\ref{sec:mergerflyby}, for which we want to measure the
distributions of the orbital parameters. The final distributions are
expected to depend on both the mass of the satellite and on the
initial distribution function. However, Fig.~\ref{fig:e2blfin}
suggests that, in particular for mergers, the location in the
$\Etwob$-$L$ space at $t=\tfin$ is not very different for
$\musat=0.001$ and $\musat=0.005$. Moreover, there is not compelling
evidence that the final distributions of $\Etwob$ and $L$ are
different in the isotropic and radially anisotropic cases (see
Appendix~\ref{app:cumul}). Therefore, in order to have a better
statistics, hereafter we consider the distributions of the final
orbital parameters of all bona-fide mergers, putting together all
families of models (I1, I5, A1 and A5).

\subsubsection{Distributions of specific energy and angular momentum}

The distributions of final $\Etwob$ and $L$ is shown in
Fig.~\ref{fig:e2blhisto}: most orbits are moderately bound
($-1\lesssim\Etwob/\sigmacen^2\lesssim 0$) and almost tangential
($L/\Lmax\gtrsim0.7$). The distribution of $\Etwob/\sigmacen^2$ is
well represented by a Gaussian with mean $\mu=-0.56$ and standard
deviation $\sigma=0.50$.  Given that, by
  definition, $0\leq L/\Lmax\leq 1$, we attempt to describe the
  distribution of $L/\Lmax$ with a beta distribution, a natural
  candidate for continuous random variables with compact support.
The probability density function of the beta distribution is
\begin{equation}
p(x)=\frac{x^{\alpha-1}(1-x)^{\beta-1}}{B(\alpha,\beta)},
\label{eq:betadist}
\end{equation}
where
\begin{equation}
B(\alpha,\beta)=\frac{\Gamma(\alpha)\Gamma(\beta)}{\Gamma(\alpha+\beta)}
\end{equation}
and $\Gamma$ is the gamma function. 
We find a good fit to the distribution of $L/\Lmax$ 
with $\alpha=1.54$ and $\beta=0.49$ (for these values the distribution
has\footnote{We recall that for a beta distribution
  $\sigma^2=\alpha\beta/[(\alpha+\beta)^2(\alpha+\beta+1)]$ and
  $\mu=\alpha/(\alpha+\beta)$.} mean $\mu=0.76$ and standard deviation
$\sigma=0.25$). The best-fitting Gaussian and beta distributions are
overplotted in Fig.~\ref{fig:e2blhisto}.  

\subsubsection{Distributions of eccentricity and pericentric radius}

For comparison with work on the orbital parameters of mergers between
halos in cosmological $N$-body simulations it is also useful to
consider the distribution of the orbital circularity $\eta$ and
pericentric radius $\rperitwob$ (see Section~\ref{sec:twobody}). As
$\eta$ is defined only for bound orbits, we restrict here our analysis
to the subsample of our simulations in which the encounter has $e<1$
(elliptic mergers).  The distributions of $\eta$ and $\rperitwob$ are
shown in Fig.~\ref{fig:etarperihisto}. The distribution of $\eta$
found for our central-satellite mergers has mean $\mu\simeq0.64$ and
standard deviation $\sigma\simeq0.28$, and is well represented by a
beta distribution (equation~\ref{eq:betadist}) with $\alpha=1.23$ and
$\beta=0.70$. The distribution of $\rperitwob/\rcen$ has mean
$\mu\simeq0.50$ and standard deviation $\sigma\simeq0.32$, and appears
consistent with being uniform.  

The distributions of $\eta$ and $\rperitwob$ that we find for
accretion onto CGs are remarkably different from the distributions
found for cosmological DM halos by \citet[][see
  Fig.~\ref{fig:etarperihisto}]{Wet11}.  Note that \citet{Wet11}
normalizes $\rperitwob$ to the virial radius, while here we normalize
to $\rcen$. However, the right-hand panel Fig.~\ref{fig:etarperihisto}
clearly shows that, independent of the somewhat arbitrary
normalization of the $x$-axis, the shape of the distribution of
$\rperitwob$ that we find for CGs is different from what found for DM
halos.  In particular, the distribution of $\eta$ for halos shown in
the left-hand panel of Fig.~\ref{fig:etarperihisto} has mean
$\mu\simeq 0.51$ and standard deviation $\sigma\simeq 0.22$, while the
distribution of $\rperitwob/\rcen$ for halos (right-hand panel of
Fig.~\ref{fig:etarperihisto}) has mean $\mu\simeq 0.24$ and standard
deviation $\sigma\simeq 0.22$.

\subsubsection{Distributions of $v/\vcirc$ and $|\vr|/v$}
\label{sec:vvr}

Another set of orbital parameters often used in the context of
cosmological $N$-body simulations \citep[e.g.][]{Ben05,Jia15} is
($v/\vcirc$, $|\vr|/v$) where, given a distance $r$ from the centre of
the host system (typically the virial radius), $v$ is the speed of the
satellite at $r$, $\vr$ is the radial component of the satellite
velocity at $r$ and $\vcirc$ is the host's circular velocity at $r$.
The final distributions of $v/\vcirc$ and $|\vr|/v$, measured by
taking as reference radius $\rcen$, are shown for all bona fide
mergers ($e<1.5$) of our simulations in Fig.~\ref{fig:vvr}. The
distribution of $v/\vcirc$ is fitted by a Gaussian with with mean
$\mu=1.39$ and standard deviation $\sigma=0.35$, while the
distribution of $|\vr|/v$ has $\mu\simeq 0.53$ and $\sigma\simeq
0.27$, and is well represented by a uniform distribution. Consistent
with the results of $\eta$ and $\rperitwob$, these distributions are
significantly different from the corresponding distributions measured
for infalling satellite halos (at the virial radius) in cosmological
$N$-body simulations by \citet{Jia15}, which are shown in
Fig.~\ref{fig:vvr} for comparison.  Specifically, the distributions
for cosmological halos shown in Fig.~\ref{fig:vvr} have mean
$\mu\simeq 1.24$ and standard deviation $\sigma\simeq0.15$ for
$v/\vcirc$, and $\mu\simeq0.79$ and $\sigma\simeq0.19$ for $|\vr|/v$.

\subsubsection{Comparison with cosmological halo-halo mergers}

A summary of the first and second moments of the distributions of the
orbital parameters ($\eta$, $\rperitwob/\rcen$, $v/\vcirc$ and
$|\vr|/v$) for the central-satellite mergers of our simulations and
for accretion of cosmological satellites is given in Table
~\ref{tab:dist}.  Based on the results found in this work and in
cosmological $N$-body simulations, we conclude that the probability
density functions of the orbital parameters of CG encounters are
substantially different from those of the corresponding functions for
cosmological halos. The orbits of satellites merging onto CGs tend to
be less bound and less eccentric than the orbits of cosmological
satellite halos when they cross the host's viral radius. Moreover, the
scatter in the orbital parameters tends to be larger for accretion
onto CGs than for accretion onto cosmological halos.

\section{Discussion}
\label{sec:disc}

The results of this work are based on an admittedly idealized models,
which rely on a few simplifying assumptions. Here we discuss these
assumptions and their possible implications.

\subsection{Initial distribution of satellite orbits}
\label{sec:initial_dist}

In our model we have assumed that the system of satellite galaxies is
initially in equilibrium in the gravitational potential of the host
(Section~\ref{sec:model}).  Though this is clearly an idealization,
this assumption is probably the most reasonable in the framework of
simple models that do not account explicitly for the cosmological
framework in which the host halo evolves.  As an alternative, one
would be tempted to consider as initial distributions of the orbital
parameters of the satellites the distributions measured for infalling
DM halos in cosmological $N$-body simulations. However, there are
reasons to believe that infalling cosmological satellites at their
first pericentric passage do not contribute significantly to accretion
onto the CG. This can be quantitatively shown by the following
argument. Let us consider, for instance, the distributions of the
orbital parameters measured by \citet{Jia15} for cosmological
infalling halos, which is expressed in terms of $v/\vcirc$ and
$|\vr|/v$ as measured at the virial radius $\rvir$ (see
Section~\ref{sec:vvr}). Let us take, for example, the best-fitting
distributions of $v/\vcirc$ and $|\vr|/v$ for hosts of mass
$10^{14}\Msun$, and mass ratio in the range 0.005-0.05. As the
distribution of $v/\vcirc$ is quite narrow, for simplicity we fix
$v/\vcirc=1.236$, the mean value of the distribution. For given
$|\vr|/v$ we then integrate the orbit in the gravitational potential
of an NFW halo with concentration $\rvir/\rs=5$ and we look for the
value of $|\vr|/v$ such that the satellite plunges down to $r=\rcen$
in the gravitational potential of the host. It turns out that only the
$0.6\%$ most radial orbits of the cosmological distribution (those
with $|\vr|/v \gtrsim 0.998$) have pericentre $\lesssim \rcen$ (see
also figure 2 in \citealt{Vul16a}) and would be classified as an
encounter.  Moreover, the typical eccentricity of such encounters (in
the two-body approximation) is $e\approx 10$, so virtually all of
these encounters are high-speed fly-bys that do not lead to rapid
mergers.  In the above estimate of the fraction of cosmological orbits
that reach the CG at the first pericentric passage we have neglected
dynamical friction. However, we have verified with $N$-body
simulations similar to those described in Section~\ref{sec:simu} that
the effect of dynamical friction, at least at the first pericentric
passage, is negligible even on such very radial orbits (see also
\citealt{Vul16a}).

As the satellites are typically not accreted at their first
pericentric passage, we expect that the cosmological orbital
distribution is substantially modified during the global relaxation
process of the host halo. Though it is not easy to predict in detail
the resulting distribution, there are reasons to believe that, owing
to phase mixing and violent relaxation, this distribution cannot be
far from equilibrium.

\subsection{Tidal stripping}
\label{sec:tidal}

In our model we have assumed that the satellite is rigid, so the
effect of tidal stripping is neglected. This simplifying assumption is
justified in the considered context, because the initial time of our
simulations ($t=0$) must not be interpreted as the time of accretion
of the satellite onto the host halo, but instead as a later time at
which the satellite has already evolved within the halo (consistent
with the assumption that the satellite's orbit is extracted from an
equilibrium distribution).  However, it is clear that our model is
simplistic in this respect. It is well known that the combination of
tidal stripping and dynamical friction leads to a complex interaction
of satellite and host stellar systems, which has been studied
extensively in the literature \citep[e.g.][]{Tor98,Taf03,Ang09}.  In
particular, \citet{Boy08} ran simulations similar to ours, but with
live satellites, therefore able to catch the combined effect of tidal
stripping and dynamical friction. \citet{Boy08} quantified the effect
of tidal stripping on the merging timescale (as the satellite loses
mass, the dynamical-friction timescale gets longer), but do not study
in detail the evolution of orbital parameters of rigid and live
satellites.  Our pairs of simulations with the same initial orbits but
different satellite mass (see Fig.~\ref{fig:el}) suggest that rigid
and live satellites should follow similar tracks in the
orbital-parameter space, though with different rates of evolution.
However, in detail, the evolution of the orbital parameters is
somewhat dependent on the satellite mass (see Fig.~\ref{fig:rpra}).

We note that \citet{Boy08} used the results of their simulations to
estimate the distribution of the circularity $\eta$ of the orbits of
satellites accreted by the CG, similar to what we have done in the
present work.  \citet{Boy08} find that the distribution of $\eta$
peaks at a lower value with respect to the distribution of
cosmological halos (figure 7 in their paper), seemingly in contrast
with the results of the present work (Fig.~\ref{fig:etarperihisto},
left-hand panel). In fact, the two results are not directly
comparable, because \citet{Boy08} measure $\eta$ at the time of
accretion on the host halo, while we measure $\eta$ at the time of the
encounter with the CG.  As a synthesis of the two results, we might
say that, from the point of view of the host halo, only satellites
that are initially on eccentric orbits are able to encounter the CG
(see also Section~\ref{sec:initial_dist}), but, when these satellites
merge with the CG, their orbits (modified by dynamical friction) tend
to be preferentially tangential from the point of view of the CG.

\subsection{Future of fly-bys}

In our analysis we have excluded encounters classified as fly-bys,
because they are not expected to lead to rapid mergers. However, it is
of course possible that at least some of such encounters will lead to
mergers in a cosmologically relevant time. In principle we could
follow in our simulations the evolution of fly-bys beyond the time
($\tfin$) of the first encounter. However, the simulations beyond
$\tfin$ would not be very meaningful, because the rigid satellite
would be a quite poor representation of a satellite passed through the
host's centre, and also the host's particle distribution would be
unrealistically modified by the rigid satellite orbiting through its
central regions, because of dynamical-friction heating
\citep{ElZ01,Nip04}. For these reasons, it is not easy to predict the
effect on the distribution of the orbital parameters of the fly-bys
that will end up merging. However, Figs.~\ref{fig:el} and
\ref{fig:elfin} (in which systems with $E/\Eu\gtrsim -0.7$ are all
fly-bys) suggest that at least some of these mergers might be
characterized by relatively low angular momentum.

\section{Summary and conclusions}
\label{sec:con}

In this paper we have studied the distribution of the orbital
parameters of satellite galaxies accreted by the CGs in groups and
clusters of galaxies. We have estimated the central-satellite orbital
parameters with the help of somewhat idealized $N$-body simulations in
which a massive satellite orbits in a stationary DM halo. We assume as
working hypothesis that the galaxies that will end up merging with the
CG originally belong to a system of satellites in equilibrium in the
gravitational potential of the host. Owing to dynamical friction, the
satellite loses energy and angular momentum and, in a fraction of the
simulations, it encounters the CG in a cosmologically relevant
timescale. In simulations that end up with an encounter between
central and satellite we measure the orbital parameters of the
encounter and we classify it either as a merger or as a fly-by. The
main findings of this work are the following.
\begin{itemize}
\item We confirm and strengthen previous indications that dynamical
  friction does not necessarily lead to orbit circularization: the
  effect of dynamical friction on the orbit of the decelerated object
  depends not only on the host's distribution function, but also, for
  given distribution function, on the initial orbital parameters of the
  satellite.
\item The distributions of the orbital parameters of the
  central-satellite mergers are markedly different from the
  distributions found for halo-halo mergers in cosmological
  simulations.
\item The orbits of the satellites accreted by the CGs are, compared
  to those of cosmological halo-halo encounters, on average less
  bound, because the merger occurs at the bottom of a deep potential
  well, and less eccentric, because the trajectory of the satellite,
  shrunk by dynamical friction, is likely to end up grazing the CG.
\item The scatter in the orbital parameters tends to be larger for
  accretion onto CGs than for accretion onto cosmological halos.
\item We provide fits to the distributions of the central-satellite
  merging orbital parameters that can be used to study the
  merger-driven evolution of the scaling relations of CGs.
\end{itemize}

\section*{Acknowledgements}

I would like to thank the referee, Jaime Perea, for the positive and
constructive report, and Benedetta Vulcani for helpful comments on a
draft of this paper.

\bibliography{cdf}
\bibliographystyle{mnras}

\appendix

\section{Final distributions of the orbital parameters for isotropic and anisotropic families of models}
\label{app:cumul}

%%%%%%%%%%%%%%FIG A1
\begin{figure*}
\centerline{\psfig{file=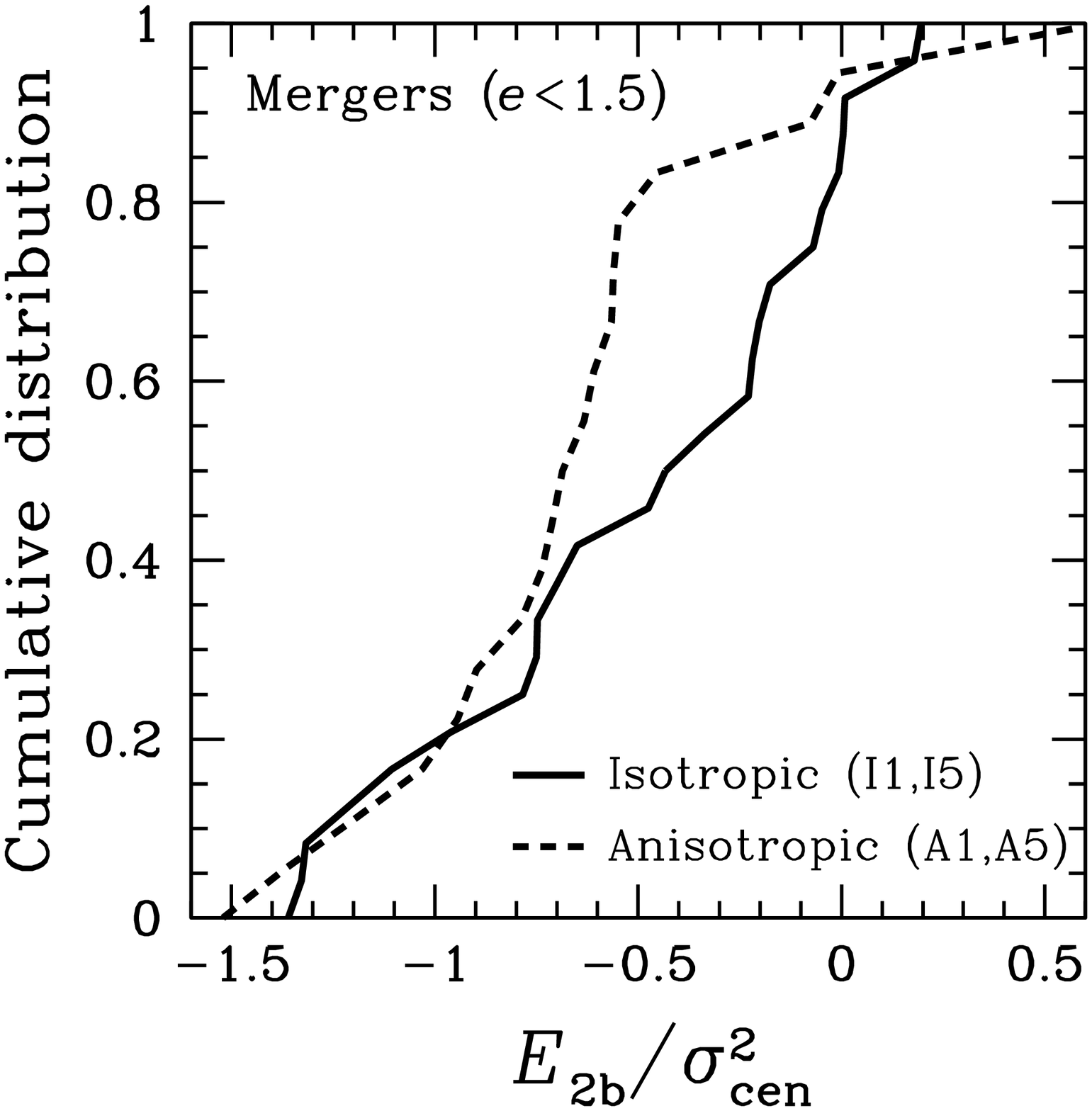,width=0.5\hsize}\psfig{file=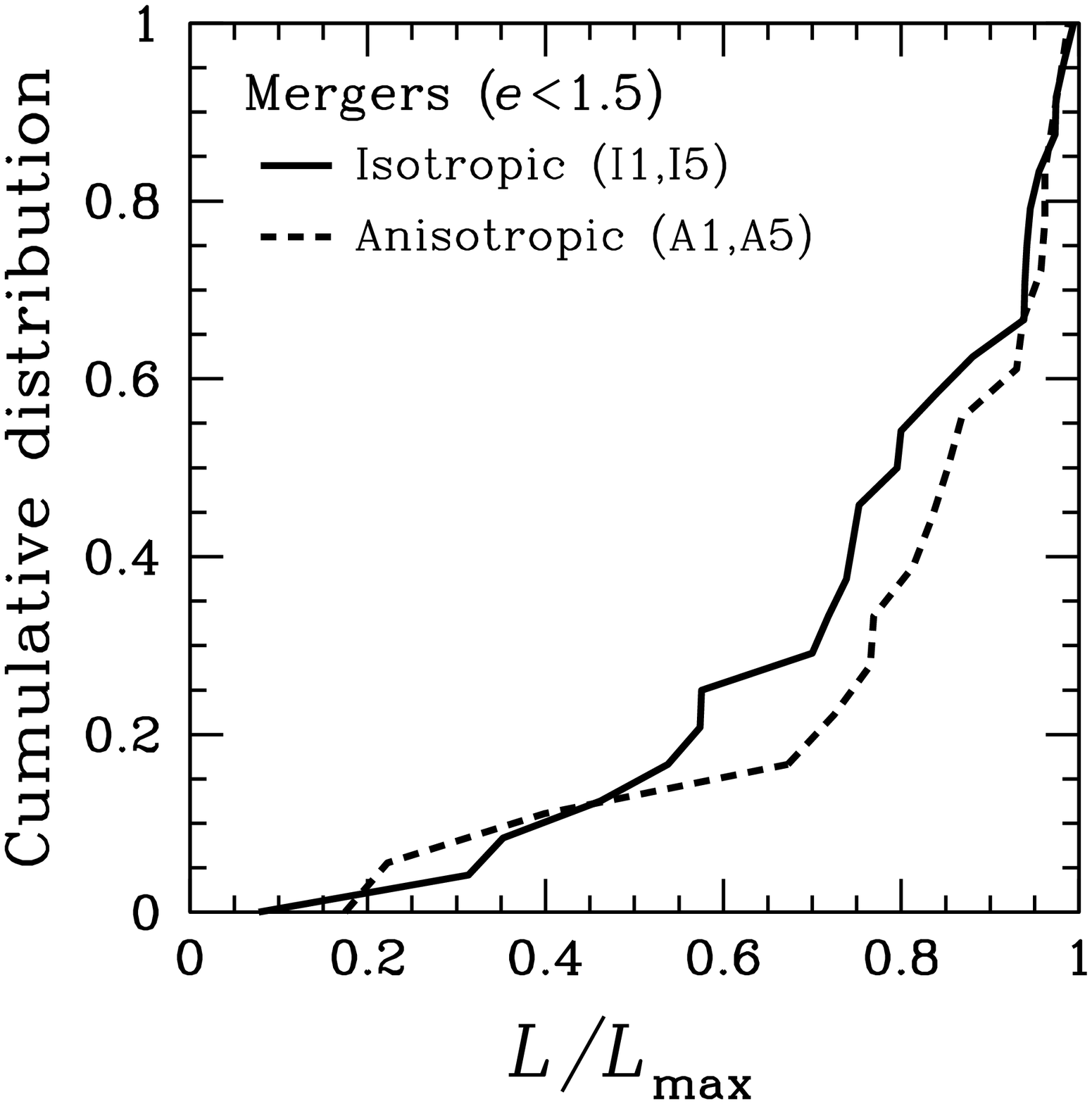,width=0.5\hsize}}
\caption{Cumulative distribution of the final specific orbital energy,
  computed in the two-body approximation ($\Etwob/\sigmacen^2$;
  left-hand panel, solid histogram), and angular-momentum modulus
  ($L/\Lmax$; right-hand panel, solid histogram) for isotropic (solid
  line) and anisotropic (dashed line) families of models. $\sigmacen$
  is the characteristic velocity dispersion of the CG and $\Lmax$ is
  defined by equation~(\ref{eq:lmax}).}
\label{fig:cumul}
\end{figure*}
%%%%%%%%%%%%%%%%%%%%%%%

In this Appendix we compare the distributions of the final orbital
parameters of central-satellite encounters in families of models with
isotropic (I1, I5) and anisotropic (A1, A5) distribution
functions. Figure~\ref{fig:cumul} plots the cumulative distributions
of specific energy $\Etwob$ (in the two-body approximation; left-hand
panel) and angular-momentum modulus $L$ (right-hand panel) for the
isotropic and anisotropic cases.  According to a Kolmogorov-Smirnov
test, the probabilities that the final distributions of $\Etwob$ and
$L$ in the isotropic and anisotropic cases are extracted from the same
parent distributions are 0.09 for $\Etwob$ and 0.63 for $L$.
Therefore, the differences between the isotropic and anisotropic cases
are more significant in the distribution of $\Etwob$ than in the
distribution of $L$, but also for $\Etwob$ the result of the
Kolmogorov-Smirnov test does not provide compelling evidence that the
final distribution of orbital energies is different in the two
families of models.

\end{document}